\documentclass[10pt,journal,epsfig]{IEEEtran}
\usepackage{graphicx,setspace}
\usepackage{subfigure,color}
\usepackage{amssymb}
\usepackage{cite,comment}
\usepackage{amsmath}
\usepackage{bm,comment}
\usepackage{algorithm}
\usepackage{algpseudocode}
\makeatletter

\newcommand{\Rmnum}[1]{\expandafter\@slowromancap\romannumeral #1@}
\makeatother
\newtheorem{fact}{Fact}
\newtheorem{proposition}{\underline{Proposition}}
\newtheorem{remark}{\underline{Remark}}

\newtheorem{lemma}{\underline{Lemma}}

\begin{document}
\title{Cellular-Connected UAV: Uplink Association, Power Control and Interference Coordination}
\author{Weidong Mei, Qingqing Wu, \IEEEmembership{Member, IEEE}, and Rui Zhang, \IEEEmembership{Fellow, IEEE}
\thanks{\scriptsize{This work has been presented in part at the IEEE Global Communications Conference, December 9-13, 2018, Abu Dhabi, United Arab Emirates\cite{cellular2018mei}.}}
\thanks{\scriptsize{W. Mei is with the NUS Graduate School for Integrative Sciences and Engineering, National University of Singapore, Singapore 119077, and also with the Department of Electrical and Computer Engineering, National University of Singapore, Singapore 117583 (e-mail: wmei@u.nus.edu).}}
\thanks{\scriptsize{Q. Wu and R. Zhang are with the Department of Electrical and Computer Engineering, National University of Singapore, Singapore 117583 (e-mails: \{elewuqq, elezhang\}@nus.edu.sg).}}}
\maketitle

\begin{comment}
To realize the large-scale usage of unmanned aerial vehicle (UAV) in the future, the emerging cellular-connected UAV communications are envisioned to be a promising solution.
\end{comment}

\begin{abstract}
The line-of-sight (LoS) air-to-ground channel brings both opportunities and challenges in cellular-connected unmanned aerial vehicle (UAV) communications. On one hand, the LoS channels make more cellular base stations (BSs) visible to a UAV as compared to the ground users, which leads to a higher macro-diversity gain for UAV-BS communications. On the other hand, they also render the UAV to impose/suffer more severe uplink/downlink interference to/from the BSs, thus requiring more sophisticated inter-cell interference coordination (ICIC) techniques with more BSs involved. In this paper, we consider the uplink transmission from a UAV to cellular BSs, under spectrum sharing with the existing ground users. To investigate the optimal ICIC design and air-ground performance trade-off, we maximize the weighted sum-rate of the UAV and existing ground users by jointly optimizing the UAV's uplink cell associations and power allocations over multiple resource blocks. However, this problem is non-convex and difficult to be solved optimally. We first propose a centralized ICIC design to obtain a locally optimal solution based on the successive convex approximation (SCA) method. As the centralized ICIC requires global information of the network and substantial information exchange among an excessively large number of BSs, we further propose a decentralized ICIC scheme of significantly lower complexity and signaling overhead for implementation, by dividing the cellular BSs into small-size clusters and exploiting the LoS macro-diversity for exchanging information between the UAV and cluster-head BSs only. Numerical results show that the proposed centralized and decentralized ICIC schemes both achieve a near-optimal performance, and draw important design insights based on practical system setups.\\
\end{abstract}

\section{Introduction}
The demand for unmanned aerial vehicles (UAVs), commonly known as drones, has been soaring globally over the recent years, due to their steadily decreasing cost and various emerging applications for e.g., aerial imaging, cargo transport, inspection, and communication platform\cite{zeng2016wireless}. As projected by the Federal Aviation Administration (FAA), the number of UAVs in civilian use, estimated at about 42,000 in 2016, will skyrocket to as many as 442,000 by 2021\cite{FAA2017}. To pave the way towards large-scale deployment of UAVs in the future, it is of paramount importance to support high-performance UAV-ground communications with ubiquitous coverage, low latency, and high reliability/throughput. This helps realize real-time command and control for UAV safe operation as well as rate-demanding payload data communication with ground users in various applications\cite{kopardekar2014unmanned}. However, at present, almost all UAVs in the market rely on the simple direct point-to-point communication with their ground pilots over the unlicensed spectrum (e.g., the industrial, scientific and medical (ISM) band at 2.4GHz). Such communication is typically of limited data rate, unreliable, insecure, vulnerable to interference, and can only operate within the visual line-of-sight (VLoS) range, thus severely limiting the future applications of UAVs.

Recently, \emph{cellular-connected UAV} has been considered as a promising new solution, by integrating UAVs into the cellular network as new aerial user equipments (UEs) served by the ground base stations (BSs). Thanks to the superior performance of today's Long Term Evolution (LTE) and future fifth-generation (5G) cellular networks, cellular-connected UAV is anticipated to achieve significant performance enhancement over the existing point-to-point UAV-ground communications over the unlicensed bands, in terms of all of reliability, coverage and throughput\cite{zeng2019cellular}. In fact, the 3rd Generation Partnership Project (3GPP) approved a new work item\cite{3GPP36777} to discuss the enhanced LTE support for aerial vehicles in early 2017. Preliminary field trials have also demonstrated that it is feasible to support the basic communication requirements for UAVs with the LTE network\cite{van2016lte,qualcom2017lte,lin2018sky}. On the other hand, thanks to the continuous improvement in UAV payload weight and communication device miniaturization, UAVs can also be utilized as aerial communication platforms (such as quasi-stationary aerial BSs/relays \cite{al2014optimal,bor2016efficient,lyu2017placement}, as well as their mobile counterparts\cite{mozaffari2017wireless,azari2018ultra,lyu2016cyclical,wu2018joint,wu2018common,wu2018uav,wu2019fundamental,zhan2011wireless,
zeng2016throughput,zhang2018joint}) to assist in terrestrial wireless networks by providing/enhancing communication services to ground UEs. This gives rise to another new research paradigm, namely \emph{UAV-aided terrestrial communication}. In this paper, we focus on the former paradigm, i.e., cellular-connected UAV communication.

Despite of the evident advantages and significant industrial efforts for cellular-connected UAV, several crucial issues need to be resolved for its efficient realization. First, how to achieve seamless and high-quality \emph{three-dimensional (3D) coverage} for both aerial and ground UEs is a challenging problem. In the current LTE network, BS antennas are usually down-tilted in order to enhance the performance of ground UEs with suppressed inter-cell interference (ICI). However, as UAVs generally fly at higher altitude than the BSs, they may be served only by the BS antenna side-lobes with weak antenna gains when integrated into the LTE network. In \cite{azari2017coexistence} and \cite{azari2018reshaping}, the coverage probability of a downlink cellular network that serves both aerial and ground users is analyzed in terms of key system parameters such as BS height, antenna pattern and UAV altitude under different BS association rules. Moreover, massive multiple-input multiple-output (MIMO) is proposed in \cite{chandhar2018massive}, where the antenna spacing for a large-size array at the BS is optimized to maximize the uplink capacity of a massive MIMO-enabled multi-UAV communication system. Second, the 3D mobility of UAVs offers additional flexibility for improving the communication performance via a \emph{communication-aware UAV trajectory design}. For example, the UAV trajectory can be flexibly designed based on the known locations of the BSs in its fly direction as well as the distribution of the ground users to ensure its communication coverage by the associated BSs and at the same time reduce the interference to/from the ground users/non-associated BSs. In \cite{zhang2019cellular}, the UAV trajectory is optimized to minimize the UAV mission completion time, subject to a quality-of-connectivity constraint with its associated BSs specified by a minimum received signal-to-noise ratio (SNR) requirement which needs to be satisfied along the UAV trajectory. Two efficient methods are proposed in \cite{zhang2019cellular} to find high-quality approximate trajectory solutions by leveraging the techniques from graph theory and convex optimization.

\begin{figure}[!t]
\centering
\includegraphics[width=3.2in]{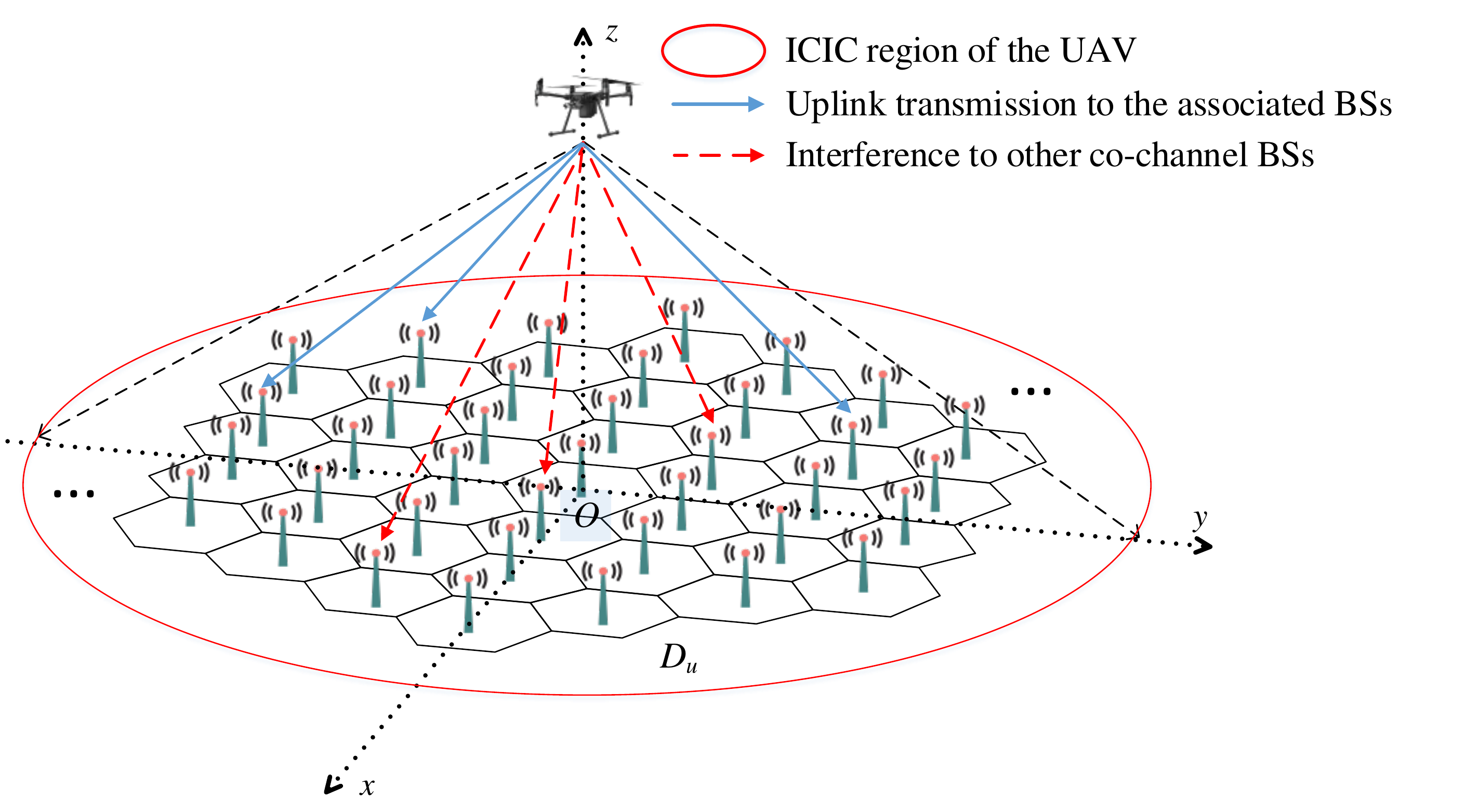}
\DeclareGraphicsExtensions.
\caption{Uplink UAV communication in a cellular network.}\label{up}
\vspace{-12pt}
\end{figure}

In this paper, we aim to address another challenging issue on how to ensure the efficient coexistence between ground and aerial UEs, via proper \emph{aerial-ground interference management}. Different from the conventional terrestrial systems, the high UAV altitude leads to unique UAV-BS line-of-sight (LoS) channels in cellular-connected UAV communication, which bring both opportunities and challenges. On one hand, the presence of LoS links leads to more reliable communication channels as compared to terrestrial channels between the ground UEs and BSs, which in general suffer from more severe path-loss, shadowing and multi-path fading. Besides, the LoS channels also make a UAV being potentially served by much more BSs at the same time, thus yielding a higher {\it macro-diversity} gain in cell associations compared to ground UEs. However, on the other hand, the dominance of LoS links also renders the UAV to impose/suffer more severe uplink/downlink interference to/from a much larger number of BSs than ground UEs. This makes the inter-cell interference coordination (ICIC) a more challenging problem to solve. Although ICIC has been extensively studied in terrestrial cellular networks (see e.g., \cite{kosta2013interference,hamza2013survey} and the references therein), such techniques may fail to mitigate the strong UAV interference and as a result lead to highly limited frequency reuse in the network after incorporating UAV UEs and hence low spectral efficiency of both ground and UAV UEs. This is because existing ICIC techniques are mainly designed to deal with the terrestrial interference to/from ground UEs, which, due to the ``unfavorable''  terrestrial channels, in fact only need to involve the coordination of at most a few cellular BSs. Whereas in cellular-connected UAV communication, due to the dominating LoS channels, a much larger ICIC region consisting of considerably more (say, tens or even hundred of) BSs is generally required (see Fig.\,\ref{up}), which incurs prohibitive complexity and overhead in practical implementation. Therefore, efficient and yet low-complexity ICIC designs are needed for enabling efficient spectrum sharing between the UAV and ground UEs in future cellular network, which, to the authors' best knowledge, have not been investigated in the literature. It is worth noting that there have been some initial studies on aerial interference mitigation in the literature\cite{zeng2019cellular,lin2018mobile,yaj2018interference,amorim2018measured}, which mainly validate the performance of existing ICIC techniques for cellular-connected UAVs via simulations or measurements, but not from an optimal design perspective.

Motivated by the above, this paper studies on the uplink ICIC design for a cellular network with co-existing UAV and ground UEs. Our goal is to mitigate the UAV's strong uplink interference to co-channel ground UEs at their associated BSs within the UAV's large ICIC region (see Fig.\,\ref{up}) and yet achieve a flexible trade-off between the performances of the UAV and ground UEs. Towards this end, we maximize the weighted sum-rate of the UAV and all ground UEs in its ICIC region by jointly optimizing the UAV's uplink cell associations and power allocations over multiple resource blocks (RBs). To tackle this problem, we first propose a centralized design by assuming that there exists a central scheduler able to collect global information from all BSs in the ICIC region and solve the design problem. As the formulated problem is non-convex, we apply an iterative successive convex approximation (SCA) algorithm to obtain a locally optimal solution. The proposed centralized design invokes the coordination of all BSs in the UAV's ICIC region and thus may incur high complexity and large delay in implementation when the number of involved BSs is too large (e.g., with small-cell BSs or high-altitude UAV). As such, we further propose a decentralized ICIC scheme of much lower complexity and signaling overhead. Specifically, we divide the cellular BSs into clusters, each with a cluster head for collecting information from its cluster BSs and exchanging information with the UAV by exploiting the LoS-induced macro-diversity. It is shown that the UAV only needs to solve an approximate convex optimization problem with the limited local information received from each cluster-head BS. The optimal solution to the approximate problem also admits a closed-form solution and thus is easy to compute. Numerical results show that both of the centralized and decentralized ICIC schemes achieve the performance close to the primal-dual based upper bound of the problem optimal value, and also greatly improve the performance over benchmark/conventional ICIC schemes. In addition, based on practical channel and system models recommended by 3GPP\cite{3GPP36777,3GPP38901}, the effects of some key system parameters (such as network loading factor, UAV altitude and antenna beamwidth) on the achievable performance are numerically analyzed and useful insights on the optimal ICIC design for cellular-connected UAV communication are obtained.

The rest of this paper is organized as follows. Sections \Rmnum{2} and \Rmnum{3} introduce the system model and formulate the problem of interest, respectively. In Section \Rmnum{4}, we propose a centralized ICIC design and solve the formulated problem by using the technique of SCA. Section \Rmnum{5} considers a decentralized ICIC design for the purpose of reducing the implementation complexity and overhead. Section \Rmnum{6} presents the simulation results to demonstrate the performance of the proposed designs. Finally, we conclude the paper in Section \Rmnum{7}.

\section{System Model}
As shown in Fig.\,\ref{up}, we consider the uplink transmission in a given subregion of the cellular network with a UAV UE and a set of ground UEs\footnote{The results of this paper can be extended to the general scenario with multiple UAV UEs in the same region with orthogonal RB allocations (e.g., via round robin or proportional fair scheme). The more challenging case with non-orthogonal RB allocations for the UAV UEs will be left as our future work.}. For simplicity, the shape of each cell is assumed to be hexagonal. For the purpose of exposition, we assume that the UAV is equipped with an isotropic antenna pointing downward\footnote{Our study can be extended to the case with directional antenna pattern of the UAV in the horizontal and vertical planes (see Section \ref{beamwidth} for details).}, while each BS employs a fixed antenna pattern (see Section \ref{sim} for details). Due to the LoS-dominated air-to-ground channel, the uplink signal from the UAV may interfere with the uplink transmissions from a large number of ground UEs using the same RBs at their associated BSs. Centered at the UAV's horizontal location on the ground, we consider there are in total $J$ BSs located in the UAV's ICIC region $D_u$, as shown in Fig.\,\ref{up}. For BSs outside this region, we assume that the signal strength from the UAV is attenuated to the level below the background noise and thus the resulted interference can be ignored. Therefore, we only need to consider the interference coordination among the $J$ BSs in the region $D_u$.\vspace{-6pt}

\subsection{Cellular Network with Ground Users Only}\label{tradICIC}
Assume that each BS $j \in {\cal J}\triangleq\{1,2,\cdots,J\}$ serves $K_j$ existing ground UEs, with $K_j \ge 1, \forall j \in \cal J$. Denote the total number of UEs in $D_u$ as $K=\sum\nolimits_{j=1}^J{K_j}$. We assume that the total number of orthogonal RBs assigned for the UAV's uplink communication is $N$, where $N < K$ usually holds in practice due to frequency reuse, among the $K$ ground UEs. For convenience, we assume that each ground UE is assigned one RB for its uplink communication, while the UAV generally needs to access multiple RBs in the uplink, due to the high-rate payload data (such as high-resolution video). To mitigate the terrestrial ICI, we consider that each BS assigns an RB to its associated UE subject to a given RB allocation criterion. Specifically, we assume that each BS checks the availability of an RB in its first $q$ tiers ($q \ge 1$) of neighboring BSs\footnote{With the considered hexagon network structure, the first $q$-tier neighboring BSs of BS $j \in \cal J$ refer to all BSs in the first $q$ rings around BS $j$.} before assigning it to a new ground UE. Let ${\cal N}_j(q)$ denote the set of the first $q$-tier neighbor BSs of BS $j$. If an RB has been occupied by a ground UE in ${\cal N}_j(q)$, BS $j$ cannot assign this RB to any new ground UE. By this means, the UEs associated with BS $j$ will not cause any interference to all cells in ${\cal N}_j(q)$. Note that when $q$ is sufficiently large, the terrestrial ICI would become negligible, thanks to the more severe path-loss and shadowing of terrestrial channels compared to the UAV-ground channels.

Accordingly, we define a set ${\cal J}(n) \subseteq \cal J$ for each given RB $n \in {\cal N}\triangleq\{1,2,\cdots,N\}$, in which $j \in {\cal J}(n)$ if RB $n$ is occupied by a ground UE in cell $j$, and as a result ${\cal J}^c(n)={\cal J}\backslash {\cal J}(n)$. Notice that under the terrestrial ICIC considered above, it must hold that ${\cal J}^c(n) \ne \emptyset, n \in \cal N$, since each BS and its first $q$-tier neighboring BSs cannot assign the same RB to their respective ground UEs simultaneously. Let $k_j(n)$ be the index of the ground UE transmitting in cell $j$ and RB $n$. Then we denote by $H_j(n)$ the channel power gain between ground UE $k_j(n)$ and its serving BS (i.e., BS $j$) in RB $n$, which in general depends on the BS antenna gain, path-loss, shadowing, and small-scale fading. The ground UE $k_j(n)$'s transmit power is assumed to be $p_{j}(n)$. Then the receive signal-to-interference-plus-noise ratio (SINR) for ground UE $k_j(n)$ at its serving BS $j$ can be expressed as
\begin{equation}
{\gamma}_{j}(n) = \frac{p_{j}(n)H_{j}(n)}{\sigma_j^2(n)},
\end{equation}
where $\sigma_j^2(n)$ is the total power of background noise and residual ICI at cell $j$ in RB $n$ (both assumed to be independently Gaussian distributed). Then the achievable sum-rate of all ground UEs in $D_u$ without the UAV's uplink transmission is given by
\begin{equation}\label{sum1}
R_g = B\sum\limits_{n = 1}^{N}{\sum\limits_{j \in {\cal J}(n)}{\log_2 (1 + {\gamma}_{j}(n))}},
\end{equation}
in bits per second (bps), with $B$ denoting the total bandwidth per RB in Hertz (Hz). For notational convenience, we denote $B=1$ Hz in the sequel of this paper, unless stated otherwise.\vspace{-12pt}

\subsection{Cellular Network with New UAV User Added}
Let ${\tilde F_j}(n)$ be the channel power gain between the UAV and BS $j$ in RB $n$. Due to the dominance of LoS propagation, we assume that the communication links from the UAV to BSs are frequency-flat over the spectrum of interest for simplicity. Thus, we have ${\tilde F_j}(n)=\tilde F_j, \forall j \in {\cal J}, n \in {\cal N}$. To exploit the LoS-induced macro-diversity, we consider a flexible cell association scheme for the UAV in which the UAV can be associated with different cells in $D_u$ over different RBs. Specifically, for all $n \in {\cal N}$, suppose that the UAV accesses an available RB $n$ in cell $j_n$ with $j_n \in {\cal J}^c(n)$ for the uplink transmission, and the UAV's transmit power is $p_n$ at RB $n$. By treating the UAV's interference as Gaussian noise for simplicity, the sum-rate of all ground UEs in RB $n$ can be expressed as
\begin{align}
R_{g,u}(n) &=\sum\limits_{j \in {\cal J}(n)} {{\log_2}\left( {1 + \frac{{p_j}(n){H_j}(n)}{\sigma_j^2(n) + {p_n}{\tilde F_j}}} \right)}\nonumber\\
&= \sum\limits_{j \in {\cal J}(n)} {\log_2}\left(1 + \frac{{{\gamma_j}(n)}}{{1 + {p_n}{F_j}(n)}}\right),\label{sum2}
\end{align}
where $F_j(n) \triangleq \tilde F_j/\sigma_j^2(n), \forall j \in {\cal J}, n \in {\cal N}$. Moreover, the achievable rate of the UAV in RB $n$ is given by
\begin{equation}\label{sum3}
R_u(n) = {\log_2}(1 + p_n{F_{j_n}(n)}).
\end{equation}

\section{Problem Formulation}\label{heuristics}
To mitigate the severe uplink interference from the UAV to its non-associated BSs and achieve a flexible performance trade-off between UAV and ground UEs, we aim to maximize the \emph{weighted} sum-rate of the UAV and all ground UEs in its ICIC region, denoted by $Q(\{j_n,p_n\}_{n \in {\cal N}})$, i.e.,
\begin{equation}\label{weight}
Q(\{j_n,p_n\}_{n \in {\cal N}})= \mu_u\sum\limits_{n \in \cal N}{R_u(n)} + \mu_g\sum\limits_{n \in \cal N}{R_{g,u}(n)},
\end{equation}
where $\mu_u \ge 0$ and $\mu_g \ge 0$ are constant weights assigned to the UAV's achievable rate and the ground UEs' sum-rate, respectively. By jointly optimizing the UAV's uplink cell associations $\{j_n\}$ and transmit power allocations $\{p_n\}$ over all RBs $n \in \cal N$, the design problem is formulated as
\begin{subequations}\label{op1}
\begin{align}
\nonumber \text{(P1)} \mathop {\max}\limits_{\{j_n,p_n\}_{n \in {\cal N}}}&\; Q(\{j_n,p_n\}_{n \in {\cal N}})\nonumber\\
\text{s.t.}\;\;&\sum\limits_{n \in {\cal N}}{p_n} \le P_{\max},\;p_n \ge 0, \forall n \in {\cal N},\label{op1a}\\
&j_n \in {\cal J}^c(n), \forall n \in {\cal N},\label{op1b}
\end{align}
\end{subequations}
where $P_{\max}$ denotes the maximum transmit power at the UAV, and constraint (\ref{op1b}) ensures that the UAV can only access RB $n$ that has not been occupied by any ground UE at its serving BS $j_n$ (but not necessarily for other non-associated BSs in ${\cal J}(n)$).
\begin{remark}\label{multiRBperUE}
It is worth noting that our system model and problem formulation are extendable to the general case with multiple RBs assigned to each ground UE. Consider, for example, a ground UE served by BS $j \in \cal J$ is assigned with $L$ RBs ($L \le N$) for its uplink communication. In this case, we can simply treat this ground UE as $L$ virtual ground UEs in the same cell, each assigned with a different RB from the $L$ RBs.
\end{remark}

\begin{remark}
By relaxing constraint (\ref{op1b}) into $j_n \in {\cal J}, \forall n \in {\cal N}$, i.e., the UAV is allowed to access each RB $n$ that has been occupied by a ground UE at its serving BS $j_n$, the network weighted sum-rate may be increased since the feasible region of (P1) is enlarged. However, according to our simulation results, constraint (\ref{op1b}) is always met at the optimality of (P1) under the practical setup that we considered. This is because constraint (\ref{op1b}) avoids causing any intra-cell interference between the UAV and the ground UEs, which can significantly degrade the UAV's achievable rate as well as the network weighted sum-rate.
\end{remark}

Note that the optimal cell association solution of (P1) can be easily obtained under any given power allocations $\{p_n\}$ based on the following lemma.
\begin{lemma}\label{solution}
The optimal cell association solution to (P1), denoted by $\{j^*_n\}$, is given by $j^*_n=\arg \mathop {\max }\limits_{j \in {\cal J}^c(n)} {F_j(n)}, \forall n \in \cal N$.
\end{lemma}
\begin{IEEEproof}
It suffices to show that for any feasible cell association solution to (P1), the corresponding objective value is no larger than $Q(\{j^*_n,p_n\})$. Suppose that $\{j^0_n\}$ is an arbitrary feasible cell association solution to (P1). It is then verified that
\begin{align}
&Q(\{j^*_n,p_n\})-Q(\{j^0_n,p_n\})\nonumber\\
=&\mu_u\sum\limits_{n \in \cal N} {{\log_2}\left(1+{p_n}{F_{j_n^*}}(n)\right)}\!-\!\mu_u\sum\limits_{n \in \cal N}{{\log }_2\left(1 + {p_n}{F_{j_n^0}}(n)\right)}.
\end{align}
As $F_{j_n^*}(n) \ge F_{j_n^0}(n)$ for any $n \in \cal N$, we must have $Q(\{j^*_n,p_n\}) \ge Q(\{j^0_n,p_n\})$.
\end{IEEEproof}

Lemma \ref{solution} implies that for each RB $n \in \cal N$, the optimal serving BS for the UAV should be the one with the maximum $F_j(n)$ among all available BSs (without any served ground UE) in the ICIC region. As such, the UAV is anticipated to achieve considerably higher macro-diversity gain in cell association as compared to ground UEs, which have far less available BSs to associate with. In addition, Lemma \ref{solution} shows that the optimal cell association solution is regardless of the UAV's transmit power allocations $\{p_n\}$, which simplifies our design.

In the sequel, we will focus on the power allocation solution to (P1) under the optimal cell association given in Lemma \ref{solution}. For ease of exposition, we define $F_u(n) \triangleq F_{j_n^*}(n), \forall n \in {\cal N}$, and (P1) is simplified as
\begin{subequations}\label{op2}
\begin{align}
\nonumber \text{(P2)} \mathop {\max}\limits_{\{p_n\}_{n \in {\cal N}}}&\; \mu_u\sum\limits_{n \in \cal N}{\log_2(1 + p_n{F_u(n)})} + \mu_g\sum\limits_{n \in \cal N}{R_{g,u}(n)}\\
\text{s.t.}\;\;&\sum\limits_{n \in {\cal N}}{p_n} \le P_{\max},\label{op2a}\\
&p_n \ge 0, \forall n \in {\cal N}.\label{op2b}
\end{align}
\end{subequations}

Note that two feasible solutions to (P2) can be easily obtained by considering an egoistic scheme and an altruistic scheme, corresponding to the optimal solutions for the two extreme cases with $\mu_g=0$ and $\mu_u=0$, respectively. Specifically, when $\mu_g=0$, the UAV only aims to maximize its own achievable rate. Obviously, the optimal power allocation in this egoistic scheme should be water-filling over all RBs, i.e.,
\begin{equation}\label{wf1}
p_n^{\text{eg}} = \left(\frac{1}{\lambda\ln 2}-\frac{1}{{F_u}(n)}\right)^+, \forall n \in \cal N,
\end{equation}
where $\left(\cdot\right)^+\triangleq\max\{\cdot,0\}$, and $\lambda$ is a constant ensuring that $\sum\nolimits_{n \in {\cal N}}{p_n^{\text{eg}}} = P_{\max}$. However, this egoistic scheme overlooks the strong interference to ground UEs and may result in significant network sum-rate loss.

On the other hand, when $\mu_u=0$, the UAV avoids causing any interference to the ground UEs in order to preserve the ground UEs' maximum sum-rate. To this end, the UAV is only permitted to transmit in the RBs that have not been occupied by any ground UEs in all cells, denoted by ${\cal N}' \triangleq \left\{n\left| n \in {\cal N}, {\cal J}(n) = \emptyset \right.\right\}$. Accordingly, if ${\cal N}'\ne \emptyset$ (i.e., when the network is not heavily loaded with ground UEs), the optimal power allocation should be water-filling over the RBs in the set ${\cal N}'$, similarly as given in (\ref{wf1}) (by replacing $n \in \cal N$ with $n \in {\cal N}'$). Otherwise, the UAV will be denied for the access to the network. The major drawback of this altruistic scheme lies in that it significantly compromises the UAV's rate performance by limiting the number of RBs available to the UAV (especially when the network is heavily loaded with ground UEs). In the next two sections, we propose more efficient solutions to (P2) than the above two simple schemes and their corresponding implementations for centralized and decentralized ICIC, respectively.

\section{Centralized ICIC}\label{central}
In this section, we solve problem (P2) by assuming that there is a central scheduler in the network. Specifically, it collects the required information from all the BSs in $D_u$, computes the power allocation (as well as cell association) solutions, and informs them to the corresponding BSs which the UAV will be associated with.\vspace{-12pt}

\subsection{Successive Convex Approximation}
Note that (P2) is a non-convex optimization problem due to its objective function, in which the second term of the ground UEs' sum-rate is not concave in the power allocation $\{p_n\}$. To efficiently solve this problem, we adopt the SCA technique to obtain a locally optimal solution. The basic idea of the SCA is to approximate the non-concave objective function as a concave one given a local point in each iteration. By iteratively solving a sequence of approximated convex problems, we can obtain a locally optimal solution to (P2).

Specifically, define $\{p_n^{(r)}\}$ as the given power allocation solution of the UAV in the $r$-th iteration. In the following, we explain how to approximate the objective function of (P2) based on the first-order Taylor approximation.
\begin{lemma}
For any given $\{p_n^{(r)}\}$, the ground UEs' sum-rate $\sum\nolimits_{n \in \cal N}{R_{g,u}(n)}$ can be lower-bounded by
\begin{equation}
\sum\limits_{n \in \cal N}{R_{g,u}(n)} \ge A^{(r)}- \sum\limits_{n \in \cal N} {B_n^{(r)}(p_n - p_n^{(r)})},\label{lb}
\end{equation}
where
\begin{align}
A^{(r)} &=\sum\limits_{n \in \cal N}{\sum\limits_{j \in {\cal J}(n)} {\log_2}\left(1 + \frac{\gamma_j(n)}{1 + p_n^{(r)}{F_j}(n)}\right)},\label{eq1}\\
B_n^{(r)} &= \sum\limits_{j \in {\cal J}(n)}{\frac{F_j(n){\gamma_j}(n)}{\ln 2(1 + p_n^{(r)}{F_j}(n) + {\gamma_j}(n))(1 + p_n^{(r)}{F_j}(n))}}.\label{eq2}
\end{align}
\end{lemma}
\begin{IEEEproof}
First, it can be shown that for each $n \in \cal N$, $R_{g,u}(n)$ given in (\ref{sum2}) is convex with respect to $p_n$. As such, the ground UEs' sum-rate $\sum\nolimits_{n \in \cal N}{R_{g,u}(n)}$ is a convex function in the UAV's power allocation $\{p_n\}$. By using the property that the first-order Taylor approximation of a convex function at any point is a global under-estimator of the convex function, we obtain the inequality (\ref{lb}).
\end{IEEEproof}

With any given local point $\{p_n^{(r)}\}$ and the lower bound given in (\ref{lb}), (P2) is approximated as the following problem in the $r$-th iteration of the SCA algorithm, i.e.,
\begin{align}
\mathop {\max}\limits_{\{p_n\}_{n \in {\cal N}}}&\; \mu_u\sum\limits_{n \in \cal N}{\log_2(1 + p_n{F_u(n)})} - \mu_g\sum\limits_{n \in \cal N} {B_n^{(r)}p_n}\label{op3}\\
\text{s.t.}\;\;&\text{(\ref{op2a}),\;(\ref{op2b})},\nonumber
\end{align}
where the constant term $\mu_g A^{(r)}+\mu_g\sum\nolimits_{n \in \cal N}{B_n^{(r)}p_n^{(r)}}$ is omitted in the objective function of (\ref{op3}) for brevity.
\begin{remark}
It is interesting to note that problem (\ref{op3}) has a price-based interpretation. The objective function of (\ref{op3}) can be viewed as a utility function for the UAV, which consists of two parts: profit (first term) and cost (second term). The cost parameter $\mu_g B_n^{(r)}$ represents the price per unit power imposed by the UAV due to its co-channel interference in RB $n$. Such interference price is iteratively updated by the SCA algorithm in order to achieve the maximum payoff (weighted sum-rate of the network).
\end{remark}

Problem (\ref{op3}) is a convex optimization problem, and thus, its optimal solution can be obtained efficiently by applying the Karush-Kuhn-Tucker (KKT) conditions (for which the details are omitted for brevity). We present the optimal solution to (\ref{op3}) in the following proposition.
\begin{proposition}\label{optPw}
The optimal solution to (\ref{op3}) is given by
\begin{equation}\label{sol1}
p_n^{(r)*}\!=\!
\begin{cases}
{\tilde p}_n^{(r)}, &\text{if}\;\sum\nolimits_{n \in \cal N} {\tilde p_n^{(r)}\!\le\!P_{\max }}\\
{\left(\frac{\mu_u}{(\mu_g B_n^{(r)}+\nu)\ln 2} - \frac{1}{F_u(n)} \right)^+}, &\text{otherwise,}
\end{cases}
\end{equation}
for all $n \in {\cal N}$, where
\[{\tilde p}_n^{(r)} \triangleq {\left(\frac{\mu_u}{\mu_gB_n^{(r)}\ln 2} - \frac{1}{F_u(n)} \right)^+},\]
and $\nu$ is a constant ensuring that $\sum\nolimits_{n \in {\cal N}}{p_n^{(r)*}} = P_{\max}$.
\end{proposition}
\begin{comment}
Similar to the process of deriving the water-filling power allocation\cite{boyd2009convex}, the optimal solution to (\ref{op3}) can be expressed as
\begin{equation}
p_n^{(r)*}=\left(\frac{1}{(\mu B_n^{(r)}+\nu)\ln 2} - \frac{1}{F_u(n)} \right)^+, \forall n \in \cal N,
\end{equation}
where $\nu$ is a dual variable corresponding to the total power constraint (\ref{op2a}). By invoking the complementary slackness in the KKT conditions, we must have $\nu(\sum\nolimits_{n \in {\cal N}}{p_n^{(r)*}}-P_{\max})=0$. Note that for (\ref{op3}), the total power constraint may not hold with equality at the optimal solution. As such, we consider two cases with $\nu=0$ and $\nu \ne 0$, respectively. Then the optimal solution given in (\ref{sol1}) can be readily obtained.
\end{comment}

The optimal power allocation in (\ref{sol1}) resembles the water-filling power allocation in (\ref{wf1}) but with the following key difference: in (\ref{sol1}) the ``water levels'' depend on the channel power gains from the UAV to all BSs $\{F_j(n)\}_{j,n}$, receive SINRs for all ground UEs $\{\gamma_j(n)\}_{j,n}$, and the rate weights $\mu_u$ and $\mu_g$, whereas their counterpart in (\ref{wf1}) is merely a constant.

After solving problem (\ref{op3}) given any local point $\{p_n^{(r)}\}$, the SCA algorithm proceeds by iteratively updating $\{p_n\}$ based on the solution to problem (\ref{op3}). Denote by $Q^{(r)}$ the objective value by the SCA algorithm in the $r$-th iteration. By applying the SCA convergence result in \cite{beck2010sequential}, it follows that a monotonic convergence is guaranteed here, i.e., $Q^{(r)} \ge Q^{(r-1)}$, $\forall r \ge 2$.

The proposed centralized ICIC scheme, which includes the above algorithm to solve (P2), is summarized in Algorithm \ref{Alg1}. For simplicity, in this paper we set the initial power allocations $\{p_n^{(1)}\}$ identical to that by the altruistic scheme and the egoistic scheme for $\mu_g \le \mu_u$ and $\mu_g > \mu_u$, respectively.
\begin{algorithm}
  \caption{Centralized ICIC Protocol}\label{Alg1}
  \begin{algorithmic}[1]
    \State The central scheduler collects the following parameters, i.e., $F_j(n), \gamma_j(n), \forall j \in {\cal J}, n \in {\cal N}$ from all BSs in $D_u$.
    \State Determine the optimal cell association solution to (P1) $\{j_n^*\}$ based on Lemma \ref{solution}.
    \State Initialize $\{p_n^{(1)}\}$. Let $r=1$.
    \State \textbf{Repeat}
    \State \quad Find the optimal solution to problem (\ref{op3}) according to Proposition \ref{optPw} as $\{p_n^{(r)*}\}$.
    \State \quad Update $p_n^{(r+1)}=p_n^{(r)*}$, $\forall n \in \cal N$.
    \State \quad Set $r=r+1$.
    \State \textbf{Until} $Q^{(r)}-Q^{(r-1)} \le \epsilon$, where $\epsilon$ is a small positive constant to control the algorithm convergence and accuracy.
    \State The central scheduler informs each serving BS in $\{j_n^*\}$ the assigned RBs $n \in \cal N$ and the UAV transmit power $\{p_n^{(r)}\}$, which are then sent to the UAV to initiate uplink data transmission.
  \end{algorithmic}
\end{algorithm}
\vspace{-12pt}

\subsection{Primal-Dual Based Upper Bound}
Though problem (P2) can only be locally optimally solved by the SCA algorithm, we can efficiently obtain an upper bound on its objective value by optimally solving its dual problem (which is convex). Note that the obtained upper bound would be tight if the strong duality holds between (P2) and its dual problem\cite{boyd2009convex}. Specifically, let $\nu \ge 0$ be the Lagrange dual variable corresponding to the total power constraint (\ref{op2a}). The partial Lagrangian of (P2) can then be expressed as
\begin{align}
{\cal L}(\{p_n\},\nu)&=\mu_u\sum\limits_{n \in \cal N} {{\log _2}(1 + {p_n}{F_u}(n))} \nonumber\\
&+ \mu_g\sum\limits_{n \in \cal N} {\sum\limits_{j \in {\cal J}(n)} {\log_2} \left( {1 + \frac{{\gamma_j}(n)}{1 + {p_n}{F_j}(n)}} \right)} \nonumber\\
&+ \nu \left(P_{\max } - \sum\limits_{n \in \cal N}{p_n}\right).\label{lag}
\end{align}

The Lagrange dual function of (P2) is then defined as
\begin{equation}\label{df}
g(\nu)= \mathop {\max}\limits_{p_n \ge 0, \forall n \in {\cal N}}\;{\cal L}(\{p_n\},\nu),
\end{equation}
which is a convex function in $\nu$. For (\ref{df}), it is easy to verify the following lemma.

\begin{lemma}
In order for the dual function $g(\nu)$ to be upper-bounded from above (i.e., $g(\nu)<\infty$), it must hold that $\nu > 0$.
\end{lemma}

Based on the lemma above, the dual problem of (P2) is given by
\begin{equation}\label{dp}
\text{(P2-D)}\;\mathop {\min}\limits_{\nu>0}\; g(\nu).
\end{equation}

Then, we can obtain an upper bound on the optimal value of (P2) by solving its dual problem (P2-D). In the following, we first solve problem (\ref{df}) to obtain $g(\nu)$ under any given $\nu>0$, and then solve (P2-D) to find the optimal $\nu$ to minimize $g(\nu)$.

Consider first the problem (\ref{df}) of maximizing the Lagrangian over $\{p_n\}$. It follows from (\ref{lag}) that problem (\ref{df}) can be decomposed into $N$ parallel subproblems, and the $n$th subproblem is given by
\begin{equation}\label{subprob}
\mathop {\max}\limits_{p_n \ge 0}\,\mu_u{\log_2}(1+{p_n}{F_u}(n))+\mu_g\!\!\!\!\!\sum\limits_{j \in {\cal J}(n)}\!\!\!\!{\log_2}\!\left(\!1\!+\!\frac{{\gamma_j}(n)}{1\!+\!p_n{F_j}(n)}\!\right)
\!-\!\nu p_n.
\end{equation}

Denote $p^D_n$ the optimal solution to (\ref{subprob}). Depending on the cardinality of ${\cal J}(n)$, we consider the following two cases to obtain $p^D_n$, respectively.

\textbf{Case 1:} If ${\cal J}(n)=\emptyset$, then the optimal solution to problem (\ref{subprob}) can be expressed as $p^D_n={\left({\frac{1}{{\nu \ln 2}} - \frac{1}{F_u(n)}} \right)^+}$ by checking the first-order derivative of the objective function of (\ref{subprob}) with respect to $p_n$.

\textbf{Case 2:} If $\lvert {\cal J}(n) \rvert \ge 1$, it is difficult to obtain the closed-form solution of $p^D_n$. Nonetheless, it can be shown that problem (\ref{subprob}) is equivalent to maximize a monotonically increasing function within a normal set\cite{phuong2003unified}. Thus, we can still obtain $p^D_n$ by applying the classical monotonic optimization technique, e.g., outer polyblock approximation (OPA) algorithm. The detailed procedures of the OPA algorithm is given in the appendix for interested readers.

After optimally solving (\ref{subprob}) for each $n \in \cal N$, the dual function $g(\nu)$ can be obtained as ${\cal L}(\{p_n^D\},\nu)$. We then optimally solve the dual problem (P2-D) to find the solution $\nu$ to minimize $g(\nu)$. As the dual function $g(\nu)$ is always convex but generally non-differentiable, problem (P2-D) can be optimally solved via applying the bisection method over $\nu$. In Section \ref{sim}, the above primal-dual based upper bound is used to evaluate the performance of the SCA algorithm numerically. If the SCA performance is sufficiently close to the upper bound, it is inferred that the SCA algorithm yields a near-optimal performance.

\section{Decentralized ICIC}
The centralized ICIC achieves locally optimal performance but requires exorbitant information exchange between the central scheduler and all involved BSs, which may incur significant overhead and large delay in the network, especially when $J$ is very large and the UAV's ICIC region dynamically changes when it moves. To reduce the implementation complexity, in this section, we propose a decentralized ICIC design by applying BS clustering and exploiting the macro-diversity thanks to the UAV-BS LoS links.

\subsection{BS Clustering}
\begin{figure}[hbtp]
\centering
\includegraphics[width=3in]{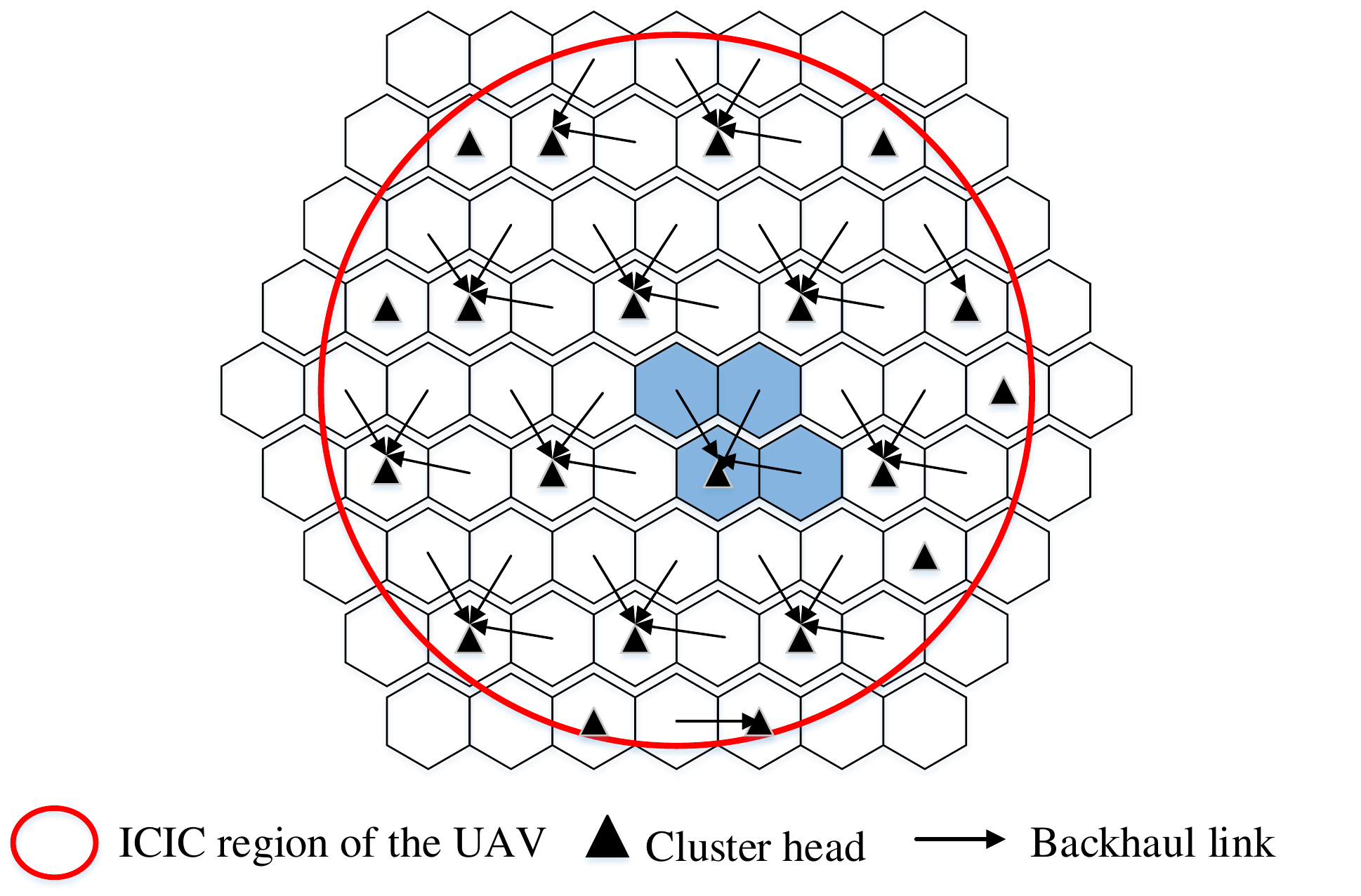}
\DeclareGraphicsExtensions.
\caption{Illustration of BS clustering with four BSs per cluster as an example.}\label{cluster.head}
\vspace{-12pt}
\end{figure}

We divide the BSs in the whole network (subsuming the ICIC region of our interest here) into non-overlapping but intra-connected clusters, and for each cluster, one BS (assumed to have a clear LoS channel with the UAV) is appointed as the cluster head to coordinate the BSs in the same cluster. For example, the cluster head can be selected as the BS with the best channel condition with the UAV. We assume that the clustering is static and the cluster size is uniform over the whole network. For example, Fig.\,\ref{cluster.head} depicts all the BS clusters in the UAV's ICIC region $D_u$ when the number of BSs per cluster is equal to 4. Upon receiving a beacon signal from the UAV, each cluster head collects the required information from other BSs in its cluster via high-speed backhaul links with low overhead (e.g, the existing X2 interface in LTE\footnote{The X2 interface has been enabled in practice to support various network functions such as ICIC, coordinated multi-point (CoMP) processing, load management, handover and so on. Interested readers may refer to \cite{dahlman20134g,3GPP36819,3GPP36423} for more details.}). The cluster heads first process the information collected independently, and report their results to the UAV via separate downlink (data or control) channels. Then the UAV solves a simplified problem of (P1) (to be specified later) with only the limited local information from the cluster heads. Assume that the total number of clusters involved in the UAV's ICIC region is $M$. Note that if $M=1$, i.e., there only exists a single cluster, the single cluster head can play the role of central scheduler to implement the centralized ICIC proposed in Section \ref{central}. Hence, we only consider the general case with $M \ge 2$ in this section. For convenience, we number all clusters/cluster heads from 1 to $M$, where cluster $m$ is denoted by $C_m$.

\subsection{Decentralized Protocol}
Next, we first show that the SCA algorithm introduced in Section \ref{central} can be implemented in a decentralized manner with the BS clustering. To this end, the UAV needs to construct and solve problem (\ref{op3}) in an iterative fashion. Let $\tilde Q(\{p_n\})$ denote the objective function of problem (\ref{op3}), i.e., $\tilde Q(\{p_n\})=\mu_u\sum\nolimits_{n \in \cal N}{\log_2(1 + p_n{F_u(n)})} - \mu_g\sum\nolimits_{n \in \cal N} {B_n^{(r)}p_n}$. Define ${{\cal J}_m}(n) = {\cal J}(n) \cap {C_m}$ and ${{\cal J}^c_m}(n) = {\cal J}^c({n}) \cap {C_m}, \forall n \in {\cal N}, m \in \cal M$, where ${\cal M}=\{1,2,\cdots,M\}$. According to (\ref{eq2}), the function $\tilde Q(\{p_n\})$ can be explicitly written as
\begin{align}
&\mu_u\sum\limits_{n \in \cal N}{\log_2}(1+p_n{F_u(n)})-\mu_g\sum\limits_{n \in \cal N}{p_n\sum\limits_{m \in \cal M}{\sum\limits_{j \in {{\cal J}_m}(n)}\!\!\!B_{j,n}^{(r)}}}\nonumber\\
=&\mu_u\sum\limits_{n \in \cal N}{\log_2}(1+p_n{F_u(n)})-\mu_g\sum\limits_{n \in \cal N}{p_n\sum\limits_{m \in \cal M} {V_{m,n}^{(r)}}},\label{eq3}
\end{align}
where $B_{j,n}^{(r)}\triangleq \frac{{F_j}(n){\gamma_j}(n)}{\ln 2(1 + p_n^{(r)}{F_j}(n) + {\gamma_j}(n))(1 + p_n^{(r)}{F_j}(n))}$ and $V_{m,n}^{(r)}\triangleq\sum\nolimits_{j \in {{\cal J}_m}(n)} B_{j,n}^{(r)}$. Moreover, based on Lemma \ref{solution}, $F_u(n)$ in (\ref{eq3}) can be rewritten as
\begin{equation}\label{eq4}
F_u(n)=\mathop {\max }\limits_{m \in {\cal M}, j \in {\cal J}_m^c(n)} {F_j}(n)=\mathop {\max }\limits_{m \in \cal M}W_{m,n},
\end{equation}
where $W_{m,n}\triangleq\mathop {\max }\limits_{j \in {\cal J}_m^c(n)} {F_j}(n)$. From (\ref{eq3}) and (\ref{eq4}), we have
\begin{equation}\label{eq5}
\tilde Q(\{p_n\})\!=\!\mu_u\!\sum\limits_{n \in \cal N}\!\log_2(1+p_n\!\mathop {\max }\limits_{m \in \cal M}\!{W_{m,n}})\!-\!\mu_g\!\sum\limits_{n \in \cal N}\!{p_n\!\!\sum\limits_{m \in \cal M}\!\!{V_{m,n}^{(r)}}}.
\end{equation}
Thus, in order to construct problem (\ref{op3}) in the $r$-th iteration, the UAV only needs to know $V_{m,n}^{(r)}$ and $W_{m,n}$, $\forall m \in {\cal M}, n \in \cal N$, which can be reported by each cluster head $m$. Specifically, the UAV first broadcasts a beacon signal to inform all BSs the current power allocation $\{p_n^{(r)}\}$ in the $r$-th iteration. Upon receiving the beacon signal, each BS $j \in \cal J$ measures the channel power gain ${F_j}(n)$ and calculates the parameter $B_{j,n}^{(r)}$ in each RB $n \in \cal N$. For each cluster head $m$, the value of $V_{m,n}^{(r)}$ can be obtained by collecting the parameters $B_{j,n}^{(r)}$ from the BSs in $C_m$ and summing them up. For the cells in ${{\cal J}^c_m}(n)$, the value of $B_{j,n}^{(r)}$ can be set to zero for convenience. On the other hand, the value of $W_{m,n}$ can be obtained by collecting the parameters ${F_j}(n)$ from the BSs in $C_m$ and taking the maximum by the cluster head of $C_m$. Similarly, for the cells in ${{\cal J}_m}(n)$, the value of ${F_j}(n)$ can be set to zero.

Hence, the centralized SCA algorithm in Algorithm \ref{Alg1} can be implemented in the following decentralized way. To start with, the UAV broadcasts a beacon signal to inform all BSs the initial power allocation $\{p_n^{(1)}\}$. Then each cluster head $m \in \cal M$ reports $2N$ parameters to the UAV in the first iteration, i.e., $V_{m,n}^{(1)}$ and $W_{m,n}$ for all $n \in \cal N$. Next, the UAV determines the optimal serving cluster head in each RB, given by $m_n^*=\arg \mathop {\max }\limits_{m \in \cal M} W_{m,n}, \forall n \in \cal N$. The optimal serving BS in RB $n$ can be found at cluster head $m_n^*$, i.e., \[j_n^*=\arg \mathop {\max }\limits_{j \in C_{m_n^*}} F_j(n).\] In addition, the UAV broadcasts the updated power allocation $\{p_n^{(2)}\}$ to the ground BSs, which can be obtained by replacing $F_u(n)$ and $B_n^{(r)}$ in (\ref{sol1}) with $\mathop {\max }\limits_{m \in \cal M}W_{m,n}$ and $\sum\nolimits_{m \in \cal M} {V_{m,n}^{(1)}}$, respectively. In the subsequent $r$-th ($r \ge 2$) iteration, each cluster head $m \in \cal M$ only needs to report $N$ parameters to the UAV, i.e., $V_{m,n}^{(r)}$ for all $n \in \cal N$, and then the UAV broadcasts the updated power allocation $\{p_n^{(r+1)}\}$ to ground BSs. The information exchange between the UAV and the cluster-head BSs proceeds until the convergence of SCA.

By this means, a locally optimal solution to (P2) can be obtained at the UAV in a decentralized manner. However, as full implementation of the SCA algorithm requires multiple information exchanges between the UAV and the cluster-head BSs, we consider a simple one-round SCA in this paper, i.e., the power allocation is only updated once at the UAV. In addition, we consider that the UAV sets the initial power allocation as $p_n^{(1)}=0, \forall n \in \cal N$ for reducing the computational burden at all BSs. As a result, each cluster head $m \in \cal M$ should report the following $2N$ simplified parameters only, i.e.,
\begin{equation}\label{eq6}
\begin{split}
&V_{m,n}=\sum\limits_{j \in {{\cal J}_m}(n)}B_{j,n},\\
&W_{m,n}=\mathop {\max }\limits_{j \in {\cal J}_m^c(n)} {F_j}(n),
\end{split}
\end{equation}
where $B_{j,n}\triangleq \frac{F_j(n){\gamma _j}(n)}{\ln 2(1 + {\gamma_j}(n))}$. Essentially, the one-round SCA aims to maximize an approximate network sum-rate, determined by its first-order Taylor approximation at the point $p_n=0, \forall n \in \cal N$. As will be shown in Section \ref{sim}, the one-round SCA can achieve a performance close to the iterative SCA. Let $\{p_n^{\text{D}}\}$ denote the computed power allocation solution at the UAV. Then the UAV should only transmit in the RBs with positive transmit power, denoted by ${\cal N}_d \triangleq \left\{n\left| n \in {\cal N}, p_n^{\text{D}}>0 \right.\right\}$. For each $n \in {\cal N}_d$, the UAV should report two parameters to all cluster heads, i.e., the indices of RB $n$ and the associated cluster head $m_n^*$. Thus, the total number of exchanged parameters is at most $2MN+2N$ in the proposed decentralized ICIC with one-round SCA. The above algorithm is summarized in Algorithm \ref{Alg2}.
\begin{algorithm}
  \caption{Decentralized ICIC Protocol}\label{Alg2}
  \begin{algorithmic}[1]
    \State The UAV broadcasts a beacon signal to initiate the protocol.
    \State Each BS $j$ individually computes $B_{j,n}$ and ${F_j}(n), \forall n \in \cal N$, and sends the values to its associated cluster head.
    \State Each cluster head $m$ individually computes $V_{m,n}$ and $W_{m,n}$ for all $n \in \cal N$ based on (\ref{eq6}), and sends the values to the UAV.
    \State The UAV determines the optimal serving cluster head $m_n^*$ in each RB $n \in \cal N$ as $m_n^*=\arg \mathop {\max }\limits_{m \in \cal M} W_{m,n}$.
    \State The UAV determines the power allocation $\{p_n^{\text{D}}\}$ based on Proposition \ref{optPw}, by replacing $F_u(n)$ and $B_n^{(r)}$ in (\ref{sol1}) with $\mathop {\max }\limits_{m \in \cal M}W_{m,n}$ and $\sum\nolimits_{m \in \cal M} {V_{m,n}}$, respectively.
    \State The UAV broadcasts $n$ and $m_n^*, \forall n \in {\cal N}_d$ to all cluster heads.
    \State Each cluster head $m_n^*$ informs BS $j_n^*$ in its cluster to initiate uplink data communication with the UAV in RB $n, \forall n \in {\cal N}_d$.
  \end{algorithmic}
\end{algorithm}
\vspace{-6pt}

\section{Simulation Results}\label{sim}
In this section, simulation results are provided to evaluate the performance of our proposed centralized and decentralized schemes. An orthogonal frequency-division multiple access (OFDMA) system is considered. Unless otherwise specified, the simulation settings are as follows. The tier of neighbor BSs is $q=2$ for the conventional terrestrial ICIC\footnote{We verify via simulations that the terrestrial ICI attenuates to the level below background noise with high probability under $q=2$ and the considered settings.}. The total number of RBs in the subband that the UAV is allowed to access is $N=30$. Each RB consists of 12 consecutive OFDM subcarriers, with the subcarrier spacing being 15 kHz. The total number of active UEs in the subband of interest is $K=60$. The transmit powers of all active ground UEs are assumed to be identical as $23$ dBm. The cell radius is $500$ m, and the height of BSs and UEs are set to be $H_B=25$ m and $H_{UE}=1.5$ m, respectively. The altitude of the UAV is fixed to $H=60$ m. The carrier frequency $f_c$ is at $2$ GHz, and the noise power spectrum density at the receiver is $-164$ dBm/Hz including a 10 dB noise figure. For the terrestrial channels, the path-loss and shadowing are modeled based on the urban macro (UMa) scenario in the 3GPP technical report\cite{3GPP38901}. The small-scale fading is modeled as Rayleigh fading. The BS antenna pattern is assumed to be directional in the vertical plane but omnidirectional in the horizontal plane. Specifically, we consider in this paper a BS antenna pattern synthesized by a uniform linear array (ULA) with 10 co-polarized dipole antenna elements\cite{ballanis2016antenna}. The antenna elements are placed vertically with half-wavelength spacing and electrically steered with downtilt angle $\theta_{\text{tilt}}=10$ degree. The ground UEs are all equipped with an isotropic antenna. On the other hand, the UAV-BS channels follow the probabilistic LoS/Non-LoS (NLoS) channel model based on the UMa scenario in the most recent 3GPP technical report\cite{3GPP36777}. The UAV's maximum transmit power $P_{\max}$ is set to be 23 dBm, same as that of ground UEs. We consider five tiers of cells centered at the cell underneath the UAV (named cell 1) to cover the UAV's ICIC region, and thus the total number of cells considered is $J=91$. The BS in cell 1 is assumed to be located at the origin without loss of generality. The UAV's horizontal location is fixed at ${\bm q}_u=$(150 m, 420 m) in cell 1. The ground UEs' locations are randomly generated in $D_u$.

In the simulation, the two heuristic schemes introduced in Section \ref{heuristics}, namely the egoistic and the altruistic schemes, are both included as benchmarks. In the egoistic scheme, the UAV only aims to maximize its own achievable rate without protecting the ground UEs' rate performance. In contrast, in the altruistic scheme, the UAV needs to preserve the ground UEs' maximum sum-rate (i.e., $R_g$ in (\ref{sum1})) by only transmitting in the RBs that have not been occupied by any ground UEs in all cells.\vspace{-12pt}

\subsection{Network Rate Performance versus UAV Transmit Power}
\begin{figure}
\centering
\begin{minipage}[t]{0.48\textwidth}
\centering
\includegraphics[width=3in]{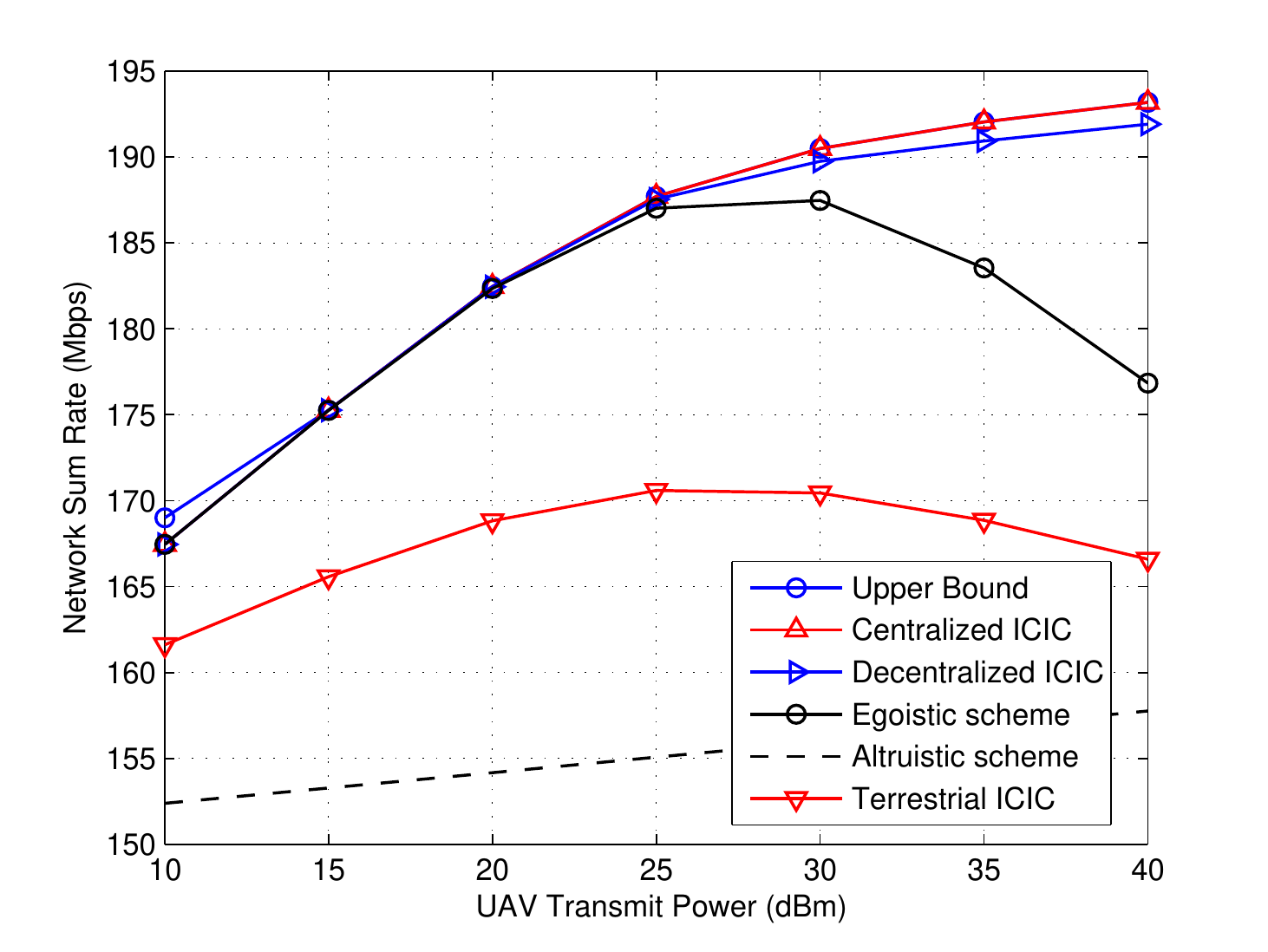}
\caption{Network sum-rate versus UAV transmit power.}\label{w.thrpt1}
\end{minipage}
\hfill
\begin{minipage}[t]{0.48\textwidth}
\centering
\includegraphics[width=3in]{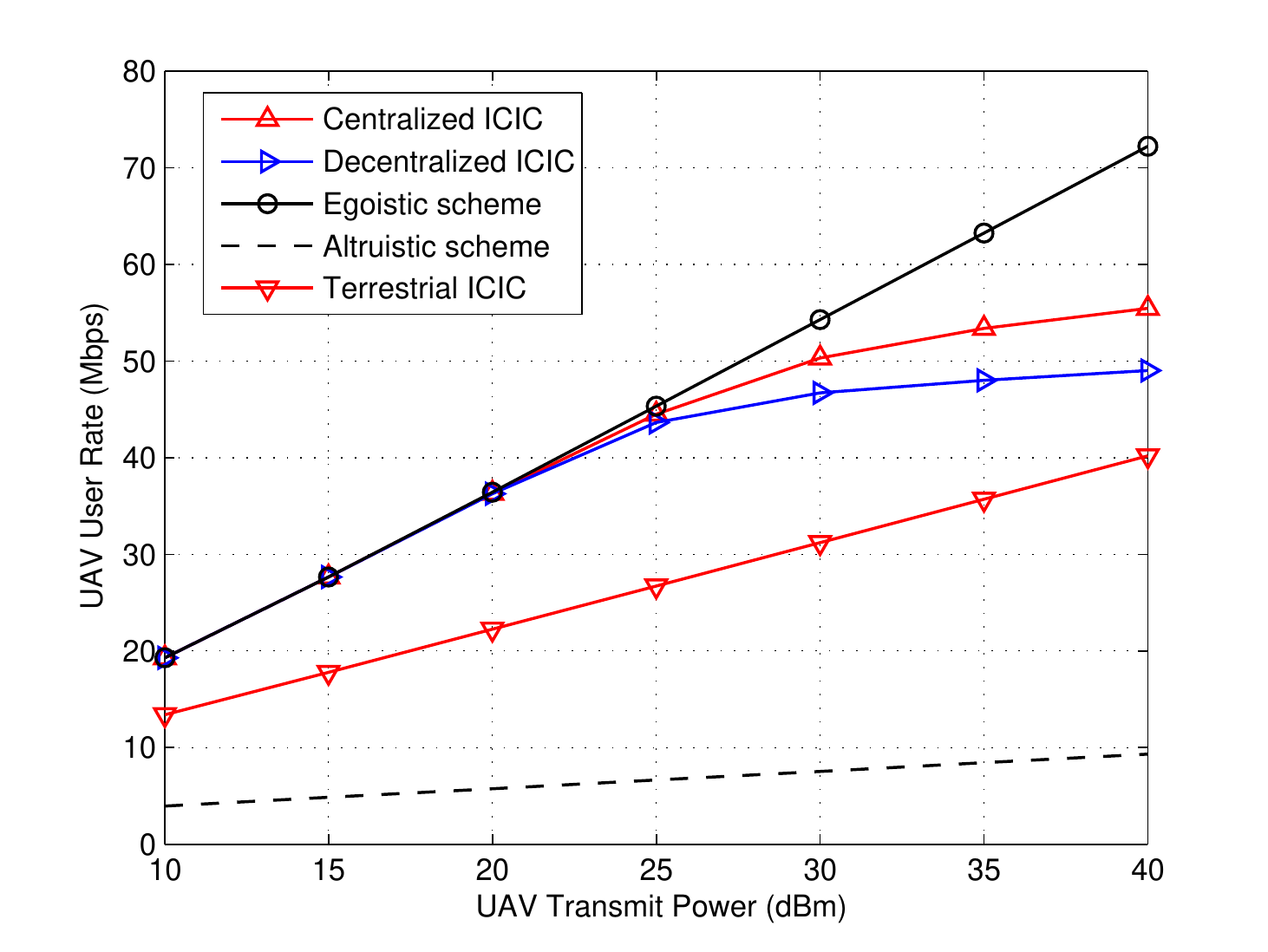}
\caption{UAV achievable rate versus UAV transmit power.}\label{w.thrpt2}
\end{minipage}
\hfill
\begin{minipage}[t]{0.48\textwidth}
\centering
\includegraphics[width=3in]{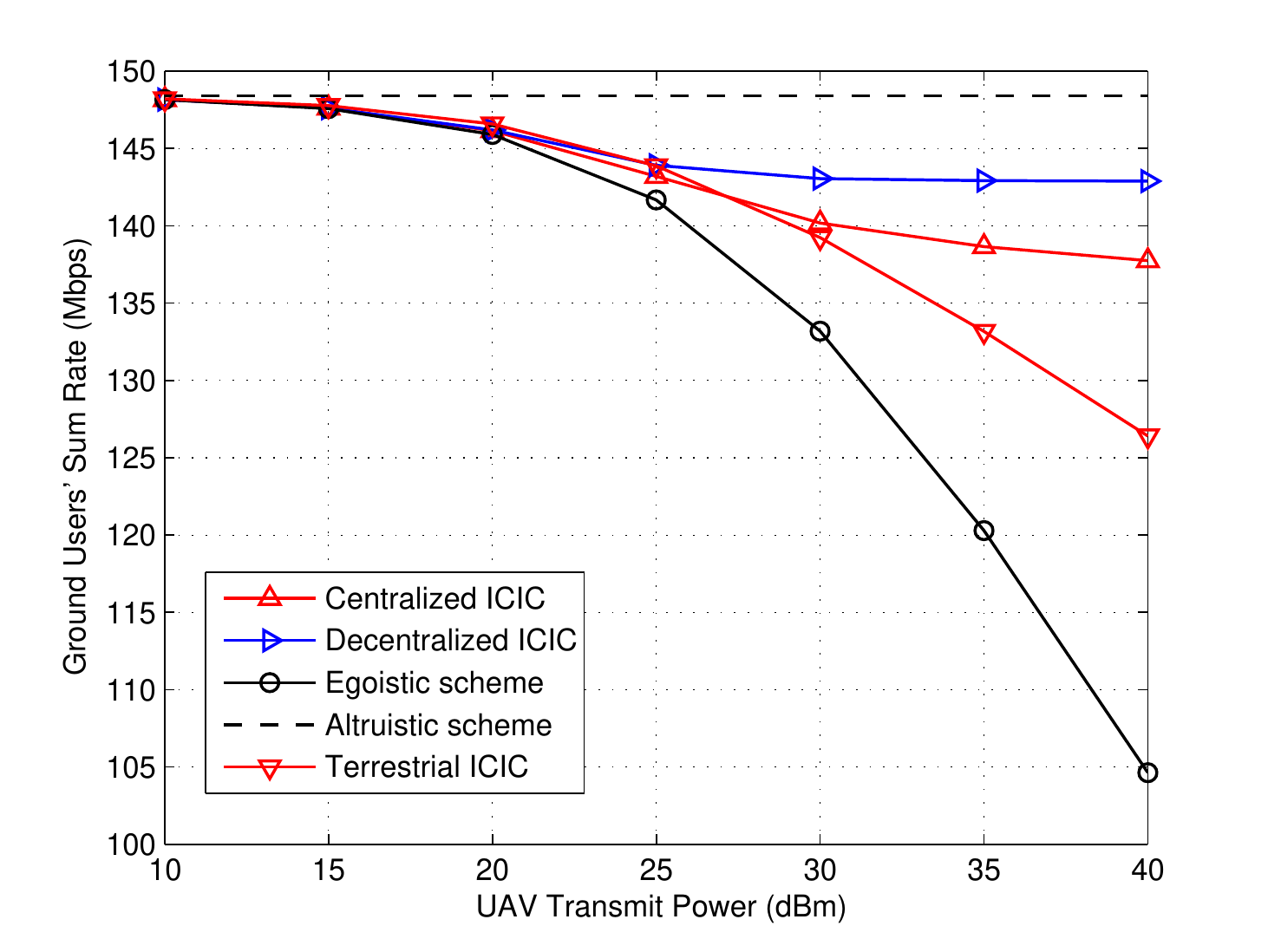}
\caption{Ground UEs' sum-rate versus UAV transmit power.}\label{w.thrpt3}
\end{minipage}
\vspace{-12pt}
\end{figure}
First, by setting $\mu_u=\mu_g=1$, Fig.\,\ref{w.thrpt1} shows the network sum-rate after integrating the UAV into the network versus the UAV's maximum transmit power $P_{\max}$, where the terrestrial ICIC described in Section \ref{tradICIC} is also included as a benchmark. In the terrestrial ICIC case, the UAV is treated as a ground UE, which simply selects a single BS with the strongest signal strength to associate with\cite{dahlman20134g}, denoted by $j_u = \arg \mathop {\max }\nolimits_{j \in \cal J}{\tilde F_j}$. Then BS $j_u$ assigns all available RBs to the UAV, subject to the RB allocation criterion introduced in Section \ref{tradICIC} with $q=2$ for ICI mitigation. The set of available RBs can be expressed as ${\cal N}^\circ \triangleq \left\{n\left| n \in {\cal N}, j_u \in {\cal J}^c(n), {\cal N}_{j_u}(q) \subseteq {\cal J}^c(n)\right.\right\}$. Since in this case the UAV causes no interference to all cells in ${\cal N}_{j_u}(q)$, we assume that the UAV applies the water-filling power control over ${\cal N}^\circ$ to maximize its achievable rate (by replacing $n \in \cal N$ in (\ref{wf1}) with $n \in {\cal N}^\circ$).
From Fig.\,\ref{w.thrpt1}, it is observed that both of the proposed centralized and decentralized ICIC designs can achieve almost the same performance as the primal-dual based upper bound, which implies that the proposed designs achieve a near-optimal performance. In addition, the gap between the centralized and decentralized ICIC designs is not large, which remains below 1.5\% over the whole range of transmit powers. It is also observed that the network sum-rate increases with the total transmit power $P_{\max}$, but at a slower rate in the high transmit power regime. This observation reveals that increasing the UAV's transmit power may not provide significant performance gain in terms of network sum-rate. This is because the rate loss of ground UEs is also increased with $P_{\max}$, and the UAV's transmit rate may not be sufficiently large to compensate for the rate loss of ground UEs. As a consequence, the UAV only consumes a fraction of its total power budget in order to maximize the network sum-rate. On the other hand, one can notice that the achievable network sum-rate by the egoistic scheme even degrades the network sum-rate in the high transmit power regime owing to the severe uplink interference caused by the UAV. In addition, the terrestrial ICIC is observed to yield a worse performance than the egoistic scheme. This is because the RB allocation criterion for terrestrial ICI avoidance limits the number of available RBs to the UAV. Moreover, it is observed that the terrestrial ICIC still degrades the network sum-rate in the high transmit power regime. The reason lies in that the size of ${\cal N}_{j_u}(q)$ is practically much smaller than that required by the UAV's ICIC region $D_u$ (see Fig.\,\ref{up}). A large number of BSs located outside ${\cal N}_{j_u}(q)$ is simply overlooked by the terrestrial ICIC and as a result they still suffer from the UAV's uplink interference. Finally, the altruistic scheme is observed to yield the worst performance of all schemes considered due to the lack of available RBs for the UAV (even fewer than with the terrestrial ICIC). The inferior performance of the benchmark/conventional schemes demonstrates the necessity of engaging more BSs for ICIC in the presence of strong UAV uplink interference, as in our proposed ICIC designs.

To further verify our observations, we plot in Fig.\,\ref{w.thrpt2} and Fig.\,\ref{w.thrpt3} the UAV achievable rate $\sum\nolimits_{n \in \cal N}{R_u(n)}$ and the sum-rate of all ground UEs $\sum\nolimits_{n \in \cal N}{R_{g,u}(n)}$, respectively. As seen from Fig.\,\ref{w.thrpt2}, the egoistic scheme gives rise to the highest UAV achievable rate over the whole range of powers. While for the two proposed ICIC schemes, the UAV achievable rate is smaller in the high transmit power regime. This result implies that from a network throughput maximization perspective, the UAV should moderately sacrifice its own rate to maximize the network sum-rate. In contrast, the altruistic scheme, as expected, yields the lowest UAV achievable rate. In addition, it is also observed that the decentralized ICIC yields lower UAV achievable rates than the centralized ICIC. This phenomenon is due to the fact that the decentralized design exaggerates the influence of UAV interference to ground UEs in the approximation, which results in more conservative transmit power allocations of the UAV. Fig.\,\ref{w.thrpt3} demonstrates that the ground UEs achieve the highest sum-rate with the altruistic scheme, and the lowest with the egoistic scheme. The terrestrial ICIC is observed to yield lower rate loss of ground UEs than the egoistic scheme, but still higher than the two proposed ICIC schemes in the high transmit power regime. Such results are consistent with those in Figs.\,\ref{w.thrpt1} and \ref{w.thrpt2}.

\begin{figure}[hbtp]
\centering
\includegraphics[width=3.2in]{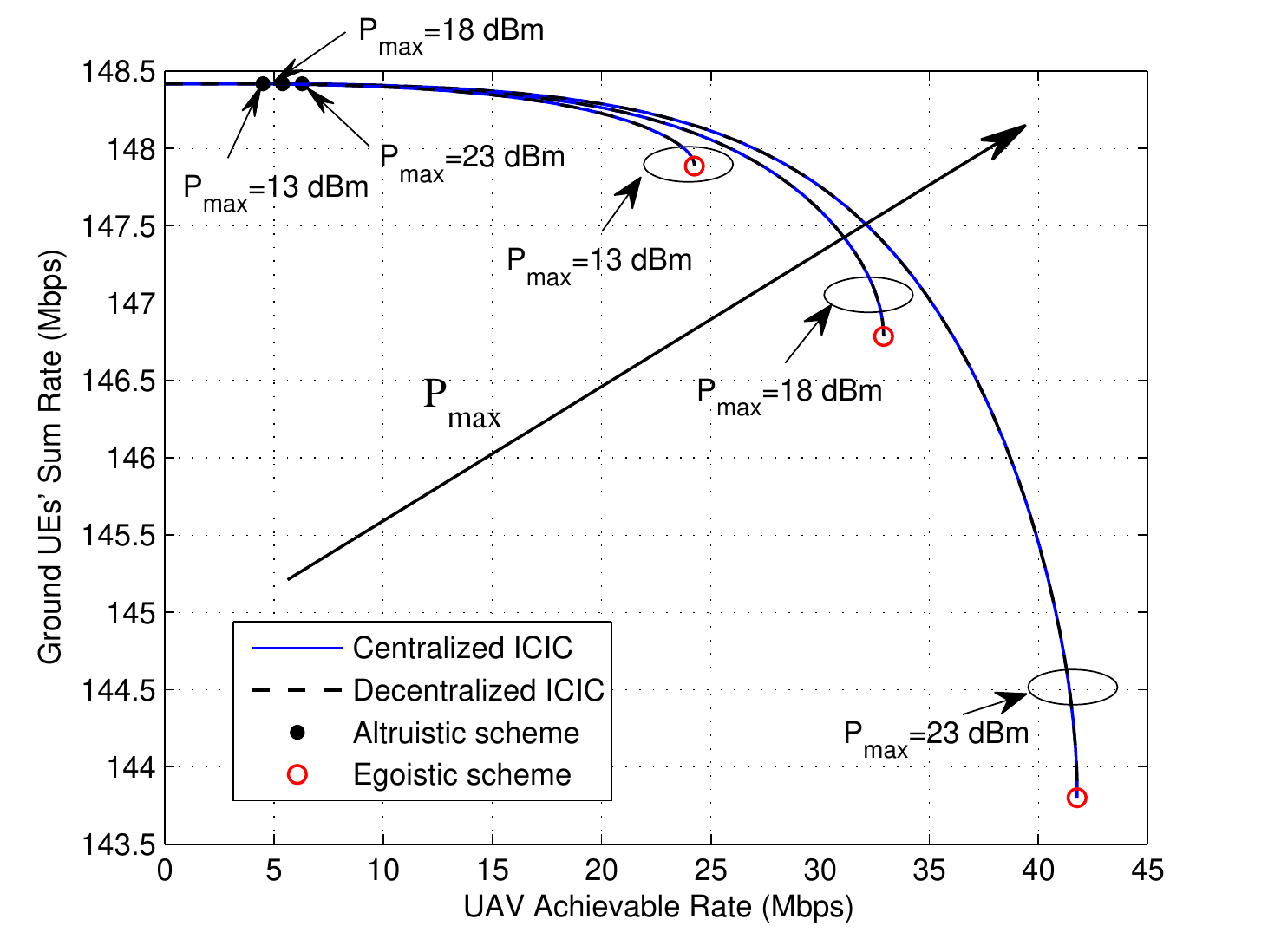}
\DeclareGraphicsExtensions.
\caption{Achievable rate region versus UAV maximum transmit power.}\label{pw.region}
\vspace{-12pt}
\end{figure}

Fig.\,\ref{pw.region} plots the achievable rate regions for the considered system with different UAV maximum transmit power $P_{\max}=13$ dBm, $18$ dBm and $23$ dBm, which characterize the trade-off between the UAV's achievable rate and the ground UEs' sum-rate by varying the ratio of $\mu_g$ to $\mu_u$. It is observed that when $P_{\max}$ is increased from $13$ dBm to $23$ dBm, the achievable rate region is also enlarged due to the increasing maximum achievable rate of the UAV. However, the boundaries of the rate regions for different $P_{\max}$ values deviate from each other more significantly when the UAV's achievable rate becomes large. This result indicates that ICIC becomes more crucial when the rate demand of the UAV is high, which is usually the case as uplink UAV communication is mainly for sending high-rate payload data (such as high-resolution video) back to the ground.\vspace{-12pt}

\subsection{Network Rate Performance versus Number of Ground UEs}
\begin{figure}[hbtp]
\centering
\includegraphics[width=3.2in]{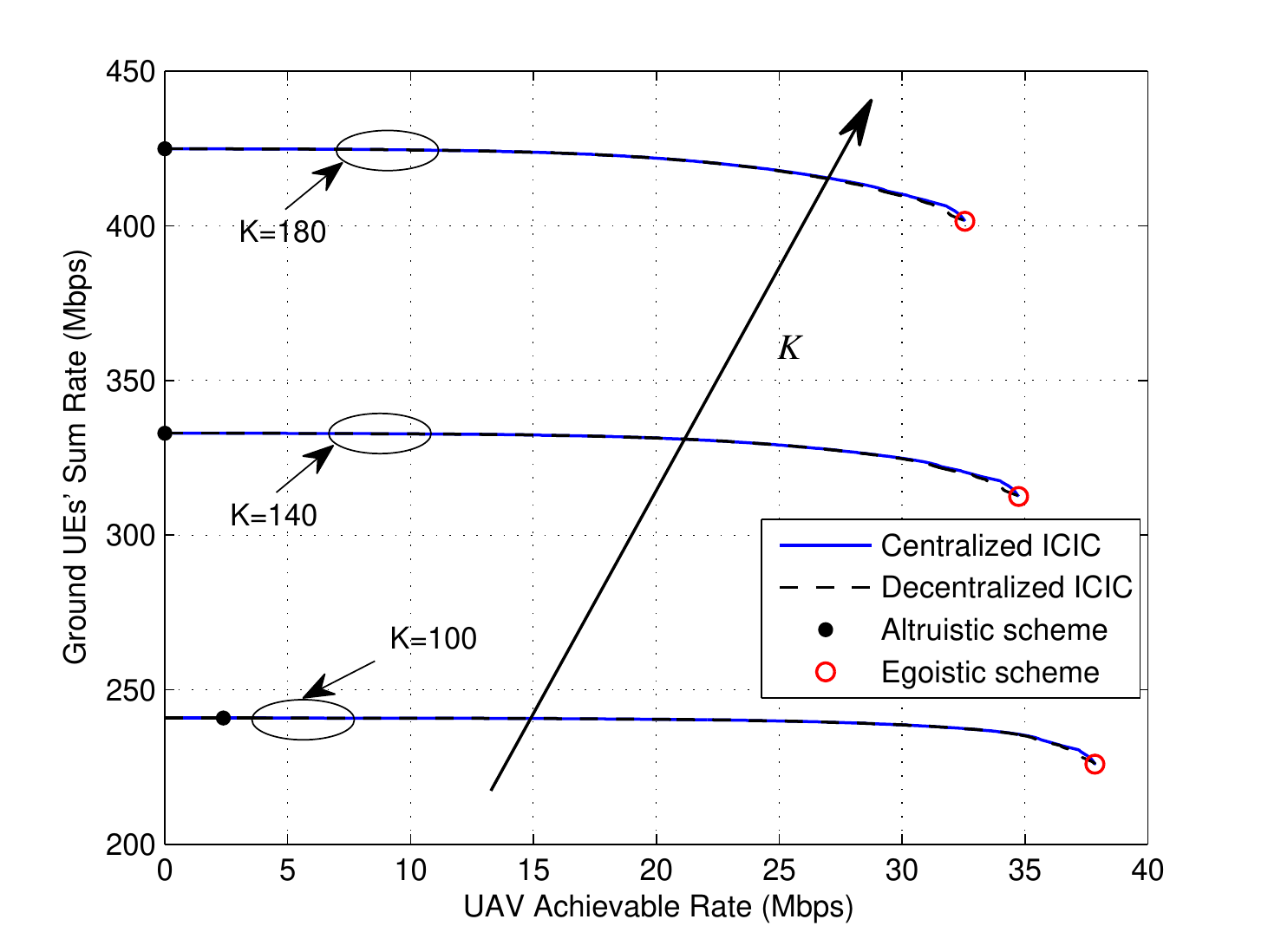}
\DeclareGraphicsExtensions.
\caption{Achievable rate region versus number of ground UEs.}\label{ds.region}
\vspace{-12pt}
\end{figure}

Fig.\,\ref{ds.region} plots the achievable rate regions for the considered system with the number of ground UEs $K=100$, $140$, and $180$, with an increasing ground traffic \emph{loading factor}. As the number of ground UEs increases, the total number of available RBs for the UAV decreases. From Fig.\,\ref{ds.region}, it is observed that with increasing $K$, the ground UEs' sum-rate is enlarged thanks to the spatial reuse of RBs. However, in contrast, the maximum UAV achievable rate by the egoistic scheme is observed to decrease with increasing $K$. This is because increasing the number of ground UEs results in higher average interference level in each RB. On the other hand, for the altruistic scheme, it is observed that the UAV achievable rate also decreases with $K$ and becomes zero with $K=140$ and $180$, i.e., the UAV is denied for the access to the network due to the lack of unoccupied RBs. The proposed ICIC designs are shown able to achieve flexible rate trade-offs between the UAV and ground UEs for different values of $K$.

\subsection{Network Rate Performance versus UAV Altitude}
In this subsection, we investigate the achievable rate region versus the UAV altitude, $H$. Although lowing UAV altitude shortens the distances from the UAV to the BSs (both associated and non-associated/interfered), and in general yields larger BS antenna (side-lobe) gains, it also increases the NLoS probability and the path-loss exponent of UAV-BS links according to \cite{3GPP36777}. Hence, there is a non-trivial relationship between the UAV altitude and the network achievable rate.

\begin{figure}[hbtp]
\centering
\includegraphics[width=3.2in]{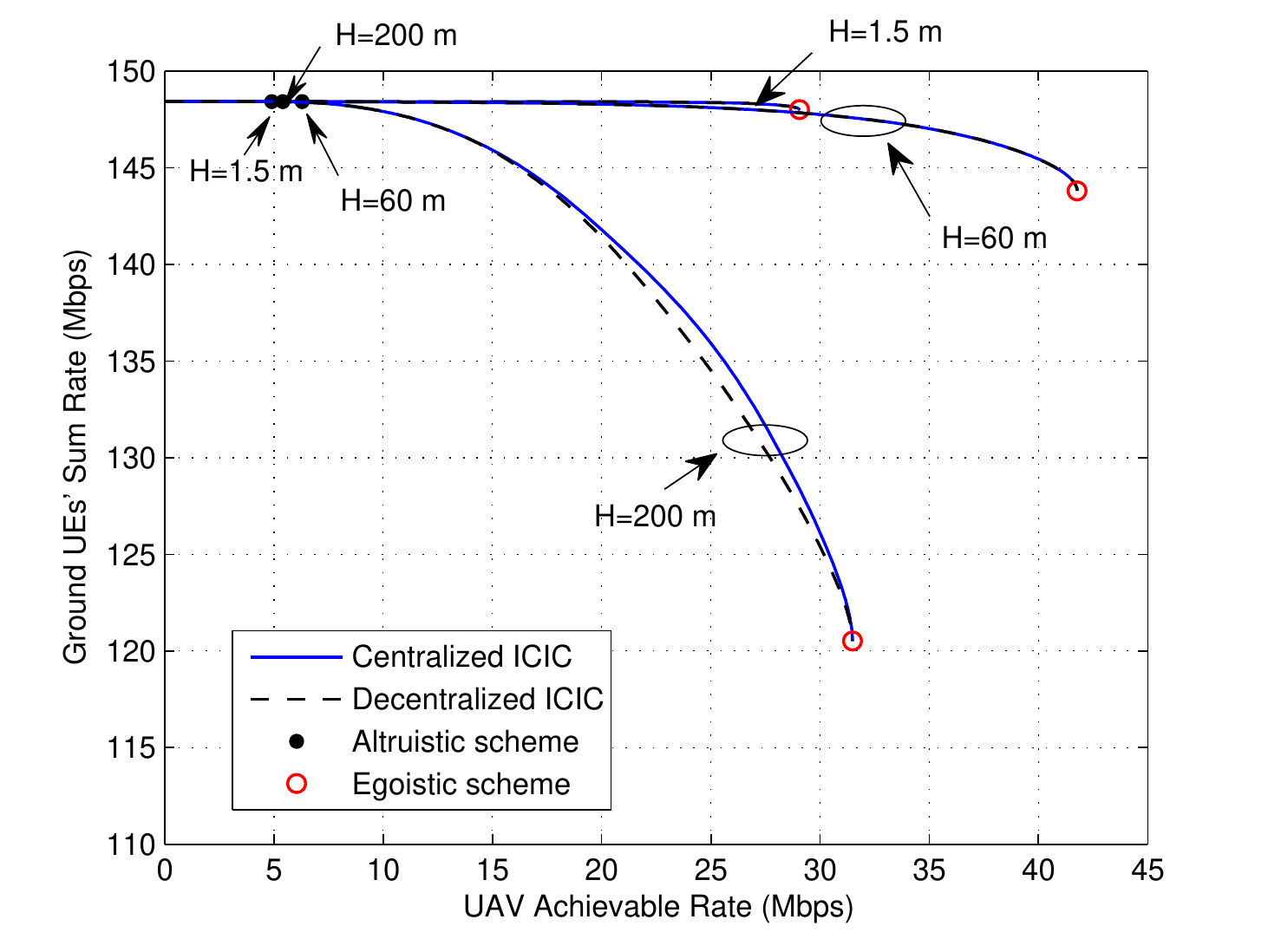}
\DeclareGraphicsExtensions.
\caption{Achievable rate region versus UAV altitude.}\label{RegionVsHt}
\vspace{-12pt}
\end{figure}

To illustrate this, we plot the achievable rate regions for the considered system with $H=1.5$ m, 60 m and 200 m in Fig.\,\ref{RegionVsHt}. Note that the case with UAV altitude 1.5 m may correspond to either a benchmark ground UE or a UAV in take-off/landing status. As seen from Fig.\,\ref{RegionVsHt}, the UAV achieves its maximum rate at a moderate altitude $H=60$ m. This is because at the lower altitude of $H=1.5$ m, the UAV achievable rate is significantly compromised by the unfavorable terrestrial channel condition and hence the lack of macro-diversity. On the other hand, at the higher altitude of $H=200$ m, the UAV will more likely fall into the antenna nulls of nearby BSs due to the down-tilted main lobe. As a result, the UAV has to be associated with more distant BSs with higher path-loss. Our simulation results show that the UAV is generally associated with 4-6 BSs at high altitude, as compared to at most 2 BSs at low-to-moderate altitude. Accordingly, the control overhead for UAV communication may be increased at high UAV altitude. A potential solution is by limiting the number of UAV's serving BSs. However, this in turn reduces the macro-diversity gain in BS association and the number of available RBs for the UAV, thus degrading the UAV's achievable rate. Moreover, from Fig.\,\ref{RegionVsHt}, it is observed that at $H=1.5$ m the ground UEs suffer the smallest rate loss when the UAV achieves its maximum rate. This is expected since the interference from the UAV is at the lowest level when $H=1.5$ m. In contrast, for $H=200$ m, the rate loss of ground UEs rapidly increases due to the increased UAV interference level. This is because the channel power gain is small between the UAV and its serving BSs at a high altitude, due to the increased distance or reduced BS antenna side-lobe gain. To achieve the maximum rate, the UAV needs to increase transmit power, which thus raises the interference level. Fig.\,\ref{RegionVsHt} reveals that from the network rate performance perspective, the UAV should operate at moderate altitude.

\subsection{Network Rate Performance versus UAV Antenna Beamwidth}\label{beamwidth}
Last, we consider that the UAV is equipped with a directional antenna with tunable beamwidth and boresight direction pointing downward. The azimuth and elevation half-power beamwidths are both assumed to be $2\Phi_u$ in degree with $\Phi_u \in (0,90^\circ)$. Specifically, the antenna gain of the UAV as seen by the BS $j \in \cal J$ can be approximately expressed as\cite{ballanis2016antenna}
\begin{equation}\label{uav.gain}
G_{u}(d_j)=
\begin{cases}
G_0/\Phi_u^2, &\text{if}\quad d_j \le r_c\\
g_0 \approx 0, &\text{otherwise,}
\end{cases}
\end{equation}
where $G_0=7500$, $d_j$ is the horizontal distance between the UAV and BS $j$ in m, and $r_c=(H-H_B)\tan\Phi_u$ is the radius of the coverage area of the UAV antenna main-lobe projected on the horizontal plane at the BS's height\footnote{By adjusting the beamwidth of the UAV's directional antenna, the size of the UAV's ICIC region can be changed accordingly.}. As seen from (\ref{uav.gain}), the antenna gain in the main lobe is reduced with increasing the antenna beamwidth. For the extreme case of $\Phi_u=90^\circ$, the UAV antenna becomes an isotropic antenna pointing downward, as considered in the previous subsections.

It is worth noting that there is in general a trade-off between maximizing the macro-diversity gain and reducing the UAV uplink interference in adjusting the antenna beamwidth of the UAV. Specifically, reducing the UAV antenna beamwidth helps reduce and even eliminate the uplink interference to ground BSs, but at the cost of UAV's own achievable rate due to the reduced number of serving BSs or macro-diversity gain. On the other hand, an increase in the UAV antenna beamwidth would cover more BSs and yield higher macro-diversity gain, but with increased uplink interference and hence rate loss of ground UEs.

\begin{figure}
\centering
\includegraphics[width=3.2in]{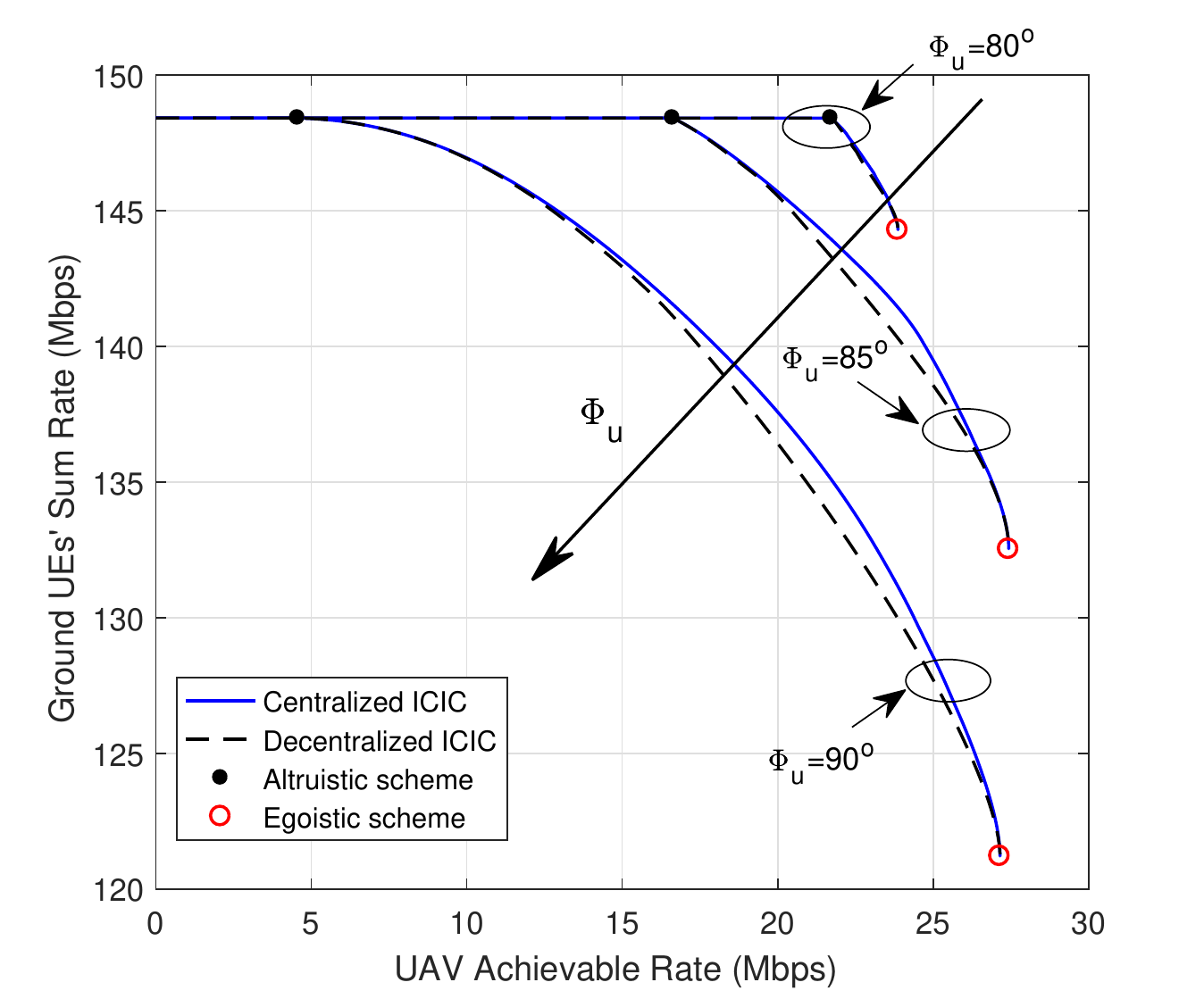}
\DeclareGraphicsExtensions.
\caption{Achievable rate region versus UAV antenna beamwidth.}\label{RegionVsTheta}
\vspace{-12pt}
\end{figure}
Fig.\,\ref{RegionVsTheta} plots the achievable rate regions for the considered system with $\Phi_u=80^\circ$, $85^\circ$ and $90^\circ$, with the UAV altitude $H$ and horizontal location ${\bm q}_u$ set to 200 m and (80 m, 100 m), respectively. It is observed that the UAV's maximum achievable rate by the egoistic scheme is significantly increased when $\Phi_u$ is increased from $80^\circ$ to $85^\circ$. This is expected since wider beamwidth brings in higher macro-diversity gain, which compensates for the loss in the UAV's antenna gain as shown in (\ref{uav.gain}). Nonetheless, the UAV's maximum achievable rate is observed to slightly decrease when $\Phi_u$ is further increased to $90^\circ$. Moreover, it is also observed that increasing the UAV antenna beamwidth results in decreasing UAV achievable rate by the altruistic scheme. This is expected since a wider beamwidth enlarges the size of ICIC region and increases the number of co-channel ground UEs. As a result, the number of unoccupied RBs is decreased, which leads to smaller rate of the UAV under the altruistic scheme. Finally, one can observe that increasing the UAV antenna beamwidth results in considerably larger rate loss of ground UEs. This is due to the rapidly enlarged ICIC region and increased number of interfered BSs. Remarkably, the achievable rate region for $\Phi_u=85^\circ$ is observed to be larger than that for $\Phi_u=90^\circ$. This demonstrates that directional antenna at the UAV can help improve the network rate performance by providing a new design degree of freedom.

\section{Conclusions}
This paper proposed new ICIC designs to mitigate the strong uplink interference due to the UAV's LoS channels with ground BSs in cellular-connected UAV communication. Specifically, the weighted sum-rate of the ground UEs and the UAV was maximized via jointly optimizing the UAV's uplink cell associations and transmit power allocations over multiple RBs. For the centralized ICIC design, it was shown that a locally optimal solution can be efficiently obtained via the SCA algorithm. To reduce the implementation complexity and overhead of the centralized ICIC, we further proposed a decentralized ICIC design, which only requires local processing within BS clusters and low-complexity signaling between the cluster-head BSs and the UAV by exploiting the UAV-ground macro-diversity. It was shown that an approximate problem can be efficiently solved in a decentralized manner, with significantly reduced overhead.

Simulation results demonstrated that the performance gap between the centralized ICIC design and the decentralized counterpart is practically small, and both achieve near-optimal rate performance. It was also demonstrated that the proposed ICIC designs are able to efficiently mitigate the air-to-ground interference and at the same time exploit the macro-diversity gain for rate enhancement, as compared to the benchmark and terrestrial ICIC schemes, especially when the UAV transmit power or the ground traffic load is high. Finally, it was shown that the network throughput is maximized by deploying the UAV at a moderate altitude and equipping the UAV with a tunable directional antenna, which help further improve the achievable rate trade-off between the UAV and ground UEs. Potential directions for future work include more advanced ICIC designs with 3D beamforming at the BS, CoMP among BSs, as well as new aerial-ground non-orthogonal multiple access (NOMA), for both UAV uplink and downlink communications.

\appendix[Solution to Problem (\ref{subprob}) via OPA Algorithm]
Due to the space limitation, we omit several important definitions that will be used in the following OPA algorithm, e.g., normal set, box, and polyblock. Readers may refer to \cite{liu2012achieving} for the detailed introduction of these notions. In order to apply the OPA algorithm for solving problem (\ref{subprob}), we need to determine an upper bound on the optimal solution $p_n^D$, as given in the following lemma.
\begin{lemma}\label{upb}
For problem (\ref{subprob}), it must hold that $p_n^D \le \hat p_n \triangleq {\left(\frac{\mu_u}{\nu \ln 2} - \frac{1}{F_u(n)}\right)^+}$.
\end{lemma}
\begin{IEEEproof}
As $p_n^D$ is the optimal solution to (\ref{subprob}), the following inequality must hold, i.e.,
\begin{align}
&\mu_u{\log_2}(1\!+\!p_n^D{F_u}(n))\!+\!\mu_g\!\!\!\!\sum\limits_{j \in {\cal J}(n)}\!\!\!{\log_2}\!\left(\!1\!+\!\frac{{\gamma_j}(n)}{1\!+\!p_n^D{F_j}(n)}\!\right)\!\!-\!\nu p_n^D \nonumber\\
\ge &\mu_u{\log _2}(1\!+\!\hat p_n{F_u}(n))\!+\!\mu_g\!\!\!\!\sum\limits_{j \in {\cal J}(n)}\!\!\!{\log_2}\!\left(\!1\!+\!\frac{{\gamma_j}(n)}{1\!+\!\hat p_n{F_j}(n)}\!\right)\!\!-\!\nu \hat p_n.\label{eqA1}
\end{align}
Notice that $\hat p_n = \arg \mathop {\max }\limits_{p_n \ge 0} \mu_u{\log_2}(1 + p_n{F_u}(n)) - \nu p_n$. Hence, it must hold that
\begin{equation}\label{eqA2}
\mu_u{\log_2}(1 + p_n^D{F_u}(n)) - \nu p_n^D \le \mu_u{\log_2}(1 + \hat p_n{F_u}(n)) - \nu \hat p_n.
\end{equation}
By combining (\ref{eqA1}) and (\ref{eqA2}), it is easy to obtain
\begin{equation}\label{eqA3}
\sum\limits_{j \in {\cal J}(n)}\!\!{\log_2}\!\left(\!1\!+\!\frac{{\gamma _j}(n)}{1\!+\!p_n^D{F_j}(n)}\!\right)\!
\ge\!\sum\limits_{j \in {\cal J}(n)}\!\!{\log_2}\!\left(\!1\!+\!\frac{{\gamma_j}(n)}{1\!+\!\hat p_n{F_j}(n)}\!\right).
\end{equation}
As the function $\sum\nolimits_{j}{\log_2} \left(1+\frac{{\gamma _j}(n)}{1 + p_n{F_j}(n)} \right)$ is monotonically decreasing with $p_n$, we then have $p_n^D \le \hat p_n$ from (\ref{eqA3}). Lemma \ref{upb} is thus proved.
\end{IEEEproof}

Next, by introducing two slack variables $z_1$ and $z_2$, it is easy to verify that the original problem (\ref{subprob}) has the same optimal solution to the following one, i.e.,
\begin{subequations}\label{opA1}
\begin{align}
\nonumber \mathop {\max}\limits_{z_1,z_2,p_n \ge 0}&\; U({\bm z}) \triangleq z_1 z_2\\
\text{s.t.}\;\;&0 \le {z_1} \le \left(1 + p_n{F_u}(n)\right)^{\mu_u},\label{opA1a}\\
&0 \le z_2 \le 2^{-\nu p_n}\prod\limits_{j \in {\cal J}(n)} \left({1+\frac{{\gamma _j}(n)}{1 + p_n{F_j}(n)}}\right)^{\mu_g},\label{opA1b}
\end{align}
\end{subequations}
where we define the vector ${\bm z}=(z_1,z_2)$.

For problem (\ref{opA1}), we have the following two facts.
\begin{fact}
The objective function of problem (\ref{opA1}) is a strictly increasing function with respect to ${\bm z}$.
\end{fact}
\begin{fact}
The feasible region of problem (\ref{opA1}), denoted by $\cal G$, is a normal set.
\end{fact}
Facts 1 and 2 imply that problem (\ref{opA1}) maximizes a strictly increasing function over a normal set. This type of problems can be solved with global optimality by using the OPA algorithm\cite{liu2012achieving}.

In the OPA algorithm, a sequence of polyblocks of shrinking sizes are iteratively constructed to approximate the feasible region $\cal G$ with the increasing accuracy for problem (\ref{opA1}). According to Lemma \ref{upb}, the polyblock can be initialized as a box $[0,{\bm z}^{(0)}]$, where
\begin{equation}\label{ini}
{\bm z}^{(0)}\!=\!(z_1^{(0)},z_2^{(0)})\!=\!\left(\left(1 + \hat p_n{F_u}(n)\right)^{\mu_u}\!,\!\prod\limits_{j \in {{\cal J}_n}} \left(1 + {\gamma_j}(n) \right)^{\mu_g}\right).
\end{equation}
The subsequent polyblocks can be successively generated by following the method presented in \cite{liu2012achieving}. In each OPA iteration, the optimal value of problem (\ref{opA1}) is found by enumeration of the vertices of a given polyblock. Let ${\bm z}^{(q)}=(z_1^{(q)},z_2^{(q)})$ denote the optimal vertex in the $q$-th iteration, i.e., \[{\bm z}^{(q)}=\arg \mathop {\max }\limits_{{\bm z} \in {\cal Z}^{(q)}} z_1 z_2,\] where ${\cal Z}^{(q)}$ represents the vertex set in the $q$-th iteration. A key step of the OPA algorithm is to compute the intersection point ${\bm r}^{(q)}$ on the Pareto boundary of the feasible region $\cal G$ with the line $\delta {\bm z}^{(q)}$. Next, we will show how to obtain such an intersection point.

In order to find $\delta$, the following optimization problem needs to be solved, i.e.,
\begin{equation}\label{opA2}
\mathop {\max}\; \delta \quad \text{s.t.}\;\;\delta {\bm z}^{(q)} \in \cal G.
\end{equation}
Problem (\ref{opA2}) is solvable via the bisection search. Specifically, given a fixed $\delta$, we need to solve the following feasibility problem, i.e.,
\begin{subequations}\label{opA3}
\begin{align}
\nonumber {\rm{find}}&\quad p_n\\
\text{s.t.}\;\;&\delta z_1^{(q)} \le \left(1 + p_n{F_u}(n)\right)^{\mu_u},\label{opA3a}\\
&\delta z_2^{(q)} \le J(p_n)\triangleq 2^{-\nu p_n}\prod\limits_{j \in {\cal J}(n)} \left({1+\frac{{\gamma _j}(n)}{1 + p_n{F_j}(n)}}\right)^{\mu_g},\label{opA3b}\\
&{p_n} \ge 0.\label{opA3c}
\end{align}
\end{subequations}
The feasibility problem can be solved efficiently as follows. First, from (\ref{opA3a}), we can obtain $p_n \ge \chi(\delta ) \triangleq \frac{{(\delta z_1^{(q)})}^{1/\mu_u} - 1}{F_u(n)}$. Let $\hat \chi(\delta)=\max \{0,\chi(\delta)\}$. Since the function $J(p_n)$ is monotonically decreasing with respect to $p_n$, problem (\ref{opA3}) is feasible if $\delta z_2^{(q)} \le J(\hat \chi(\delta))$; otherwise, it is infeasible. By updating the upper and lower bounds on $\delta$, the optimal solution to (\ref{opA2}), denoted by $\delta^{(q)}$, can be found. The intersection point ${\bm r}^{(q)}$ should be $(\delta^{(q)}z_1^{(q)},\delta^{(q)}z_2^{(q)})$. Then the vertex set is updated by replacing the point ${\bm z}^{(q)}=(z_1^{(q)},z_2^{(q)})$ with two new points $(\delta^{(q)}z_1^{(q)},z_2^{(q)})$ and $(z_1^{(q)},\delta^{(q)}z_2^{(q)})$. The above algorithm is summarized in Algorithm \ref{Alg3}.
\begin{algorithm}
  \caption{OPA Algorithm for Solving Problem (\ref{subprob})}\label{Alg3}
  \begin{algorithmic}[1]
    \State Initialize $q = 1$ and ${\cal Z}^{(1)}=\{{\bm z}^{(0)}\}$.
    \While {$\epsilon$-accuracy is not reached with $\epsilon$ denoting a small positive constant}
    \State Find the optimal vertex ${\bm z}^{(q)}$ in the set ${\cal Z}^{(q)}$ based on \[{\bm z}^{(q)}=(z_1^{(q)},z_2^{(q)})=\arg \mathop {\max }\limits_{{\bm z} \in {\cal Z}^{(q)}} z_1 z_2.\]
    \State Compute $\delta^{(q)}$ and obtain the intersection point ${\bm r}^{(q)}=(\delta^{(q)}z_1^{(q)},\delta^{(q)}z_2^{(q)})$ by solving the feasibility problem (\ref{opA3}) and utilizing the bisection search.
    \State Update the best intersection point up to the $q$-th iteration, i.e., \[{\tilde {\bm r}}^{(q)}=\arg \mathop {\max } \{U({\bm r}^{(q)}),U({\tilde {\bm r}}^{(q-1)})\}.\]
    \If {$U({\bm z}^{(q)})-U({\tilde {\bm r}}^{(q)}) \le \epsilon$}
    \State Stop and ${\tilde {\bm r}}^{(q)}$ is an $\epsilon$-optimal solution to problem (\ref{subprob}).
    \Else
    \State Update the vertex set based on \[{\cal Z}^{(q+1)}={\cal Z}^{(q)}\backslash z^{(q)} \cup \{({\delta ^{(q)}}z_1^{(q)},z_2^{(q)}),(z_1^{(q)},{\delta ^{(q)}}z_2^{(q)})\}.\]
    \EndIf
    \State Set $q=q+1$.
    \EndWhile
  \end{algorithmic}
\end{algorithm}

\bibliography{UAV_ICIC}

\begin{thebibliography}{10}
\providecommand{\url}[1]{#1}
\csname url@samestyle\endcsname
\providecommand{\newblock}{\relax}
\providecommand{\bibinfo}[2]{#2}
\providecommand{\BIBentrySTDinterwordspacing}{\spaceskip=0pt\relax}
\providecommand{\BIBentryALTinterwordstretchfactor}{4}
\providecommand{\BIBentryALTinterwordspacing}{\spaceskip=\fontdimen2\font plus
\BIBentryALTinterwordstretchfactor\fontdimen3\font minus
  \fontdimen4\font\relax}
\providecommand{\BIBforeignlanguage}[2]{{%
\expandafter\ifx\csname l@#1\endcsname\relax
\typeout{** WARNING: IEEEtran.bst: No hyphenation pattern has been}%
\typeout{** loaded for the language `#1'. Using the pattern for}%
\typeout{** the default language instead.}%
\else
\language=\csname l@#1\endcsname
\fi
#2}}
\providecommand{\BIBdecl}{\relax}
\BIBdecl

\bibitem{cellular2018mei}
W.~Mei, Q.~Wu, and R.~Zhang, ``Cellular-connected {UAV:} {U}plink association,
  power control and interference coordination,'' in \emph{Proc. {IEEE} Global
  Commun. Conf. (Globecom)}, Abu Dhabi, UAE, Dec. 2018.

\bibitem{zeng2016wireless}
Y.~Zeng, R.~Zhang, and T.~J. Lim, ``Wireless communications with unmanned
  aerial vehicles: Opportunities and challenges,'' \emph{IEEE Commun. Mag.},
  vol.~54, no.~5, pp. 36--42, May 2016.

\bibitem{FAA2017}
\BIBentryALTinterwordspacing
``{FAA} forecasts continued growth in air travel,'' 2017. [Online]. Available:
  \url{https://www.faa.gov/news/updates/?newsId=87746\&cid=TW502}
\BIBentrySTDinterwordspacing

\bibitem{kopardekar2014unmanned}
\BIBentryALTinterwordspacing
P.~H. Kopardekar, ``Unmanned aerial system ({UAS}) traffic management ({UTM}):
  Enabling low-altitude airspace and {UAS} operations,'' 2014. [Online].
  Available:
  \url{https://ntrs.nasa.gov/archive/nasa/casi.ntrs.nasa.gov/20140013436.pdf}
\BIBentrySTDinterwordspacing

\bibitem{zeng2019cellular}
Y.~Zeng, J.~Lyu, and R.~Zhang, ``Cellular-connected {UAV}: Potential,
  challenges, and promising technologies,'' \emph{{IEEE} Wireless Commun.},
  vol.~26, no.~1, pp. 120--127, Feb. 2019.

\bibitem{3GPP36777}
\BIBentryALTinterwordspacing
{3GPP-TR-36.777}, ``Study on enhanced {LTE} support for aerial vehicles,''
  2017, 3GPP technical report. [Online]. Available:
  \url{www.3gpp.org/dynareport/36777.htm}
\BIBentrySTDinterwordspacing

\bibitem{van2016lte}
B.~Van~der Bergh, A.~Chiumento, and S.~Pollin, ``{LTE} in the sky: Trading off
  propagation benefits with interference costs for aerial nodes,'' \emph{IEEE
  Commun. Mag.}, vol.~54, no.~5, pp. 44--50, May 2016.

\bibitem{qualcom2017lte}
\BIBentryALTinterwordspacing
{Qualcomm Technologies, Inc.}, ``{LTE} unmanned aircraft systems trial
  report,'' May 2017. [Online]. Available:
  \url{https://www.qualcomm.com/documents/lte-unmanned-aircraft-systems-trial-%
report}
\BIBentrySTDinterwordspacing

\bibitem{lin2018sky}
X.~Lin \emph{et~al.}, ``The sky is not the limit: {LTE} for unmanned aerial
  vehicles,'' \emph{IEEE Commun. Mag.}, vol.~56, no.~4, pp. 204--210, Apr.
  2018.

\bibitem{al2014optimal}
A.~Al-Hourani, S.~Kandeepan, and S.~Lardner, ``Optimal {LAP} altitude for
  maximum coverage,'' \emph{{IEEE} Wireless Commun. Lett.}, vol.~3, no.~6, pp.
  569--572, Dec. 2014.

\bibitem{bor2016efficient}
R.~I. Bor-Yaliniz, A.~El-Keyi, and H.~Yanikomeroglu, ``Efficient 3-{D}
  placement of an aerial base station in next generation cellular networks,''
  in \emph{Proc. {IEEE} Int. Conf. Commun. (ICC)}, Kuala Lumpur, Malaysia, May
  2016, pp. 1--5.

\bibitem{lyu2017placement}
J.~Lyu, Y.~Zeng, R.~Zhang, and T.~J. Lim, ``Placement optimization of
  {UAV}-mounted mobile base stations,'' \emph{{IEEE} Commun. Lett.}, vol.~21,
  no.~3, pp. 604--607, Mar. 2017.

\bibitem{mozaffari2017wireless}
M.~Mozaffari, W.~Saad, M.~Bennis, and M.~Debbah, ``Wireless communication using
  unmanned aerial vehicles ({UAV}s): Optimal transport theory for hover time
  optimization,'' \emph{{IEEE} Trans. Wireless Commun.}, vol.~16, no.~12, pp.
  8052--8066, Dec. 2017.

\bibitem{azari2018ultra}
M.~M. Azari, F.~Rosas, K.-C. Chen, and S.~Pollin, ``Ultra reliable {UAV}
  communication using altitude and cooperation diversity,'' \emph{{IEEE} Trans.
  Commun.}, vol.~66, no.~1, pp. 330--344, Jan. 2018.

\bibitem{lyu2016cyclical}
J.~Lyu, Y.~Zeng, and R.~Zhang, ``Cyclical multiple access in {UAV}-aided
  communications: A throughput-delay tradeoff,'' \emph{{IEEE} Wireless Commun.
  Lett.}, vol.~5, no.~6, pp. 600--603, Dec. 2016.

\bibitem{wu2018joint}
Q.~Wu, Y.~Zeng, and R.~Zhang, ``Joint trajectory and communication design for
  multi-{UAV} enabled wireless networks,'' \emph{{IEEE} Trans. Wireless
  Commun.}, vol.~17, no.~3, pp. 2109--2121, Mar. 2018.

\bibitem{wu2018common}
Q.~Wu and R.~Zhang, ``Common throughput maximization in {UAV}-enabled {OFDMA}
  systems with delay consideration,'' \emph{{IEEE} Trans. Commun.}, vol.~66,
  no.~12, pp. 6614--6627, Dec. 2018.

\bibitem{wu2018uav}
Q.~Wu, J.~Xu, and R.~Zhang, ``Capacity characterization of {UAV}-enabled
  two-user broadcast channel,'' \emph{{IEEE} J. Sel. Areas Commun.}, vol.~36,
  no.~9, pp. 1955--1971, Sep. 2018.

\bibitem{wu2019fundamental}
Q.~Wu, L.~Liu, and R.~Zhang, ``Fundamental trade-offs in communication and
  trajectory design for {UAV}-enabled wireless network,'' \emph{{IEEE} Wireless
  Commun.}, vol.~26, no.~1, pp. 36--44, Feb. 2019.

\bibitem{zhan2011wireless}
P.~Zhan, K.~Yu, and A.~L. Swindlehurst, ``Wireless relay communications with
  unmanned aerial vehicles: Performance and optimization,'' \emph{IEEE Trans.
  Aerosp. Electro. Syst.}, vol.~47, no.~3, pp. 2068--2085, Jul. 2011.

\bibitem{zeng2016throughput}
Y.~Zeng, R.~Zhang, and T.~J. Lim, ``Throughput maximization for {UAV}-enabled
  mobile relaying systems,'' \emph{{IEEE} Trans. Commun.}, vol.~64, no.~12, pp.
  4983--4996, Dec. 2016.

\bibitem{zhang2018joint}
S.~Zhang, H.~Zhang, Q.~He, K.~Bian, and L.~Song, ``Joint trajectory and power
  optimization for {UAV} relay networks,'' \emph{{IEEE} Commun. Lett.},
  vol.~22, no.~1, pp. 161--164, Jan. 2018.

\bibitem{azari2017coexistence}
M.~M. Azari, F.~Rosas, A.~Chiumento, and S.~Pollin, ``Coexistence of
  terrestrial and aerial users in cellular networks,'' in \emph{Proc. IEEE
  Global Commun. Conf. (Globecom) Wkshps.}, Singapore, Dec. 2017.

\bibitem{azari2018reshaping}
M.~M. Azari, F.~Rosas, and S.~Pollin, ``Reshaping cellular networks for the
  sky: The major factors and feasibility,'' in \emph{Proc. {IEEE} Int. Conf.
  Commun. (ICC)}, Kansas City, MO, USA, May 2018.

\bibitem{chandhar2018massive}
P.~Chandhar, D.~Danev, and E.~G. Larsson, ``Massive {MIMO} for communications
  with drone swarms,'' \emph{{IEEE} Trans. Wireless Commun.}, vol.~17, no.~3,
  pp. 1604--1629, Mar. 2018.

\bibitem{zhang2019cellular}
S.~Zhang, Y.~Zeng, and R.~Zhang, ``Cellular-enabled {UAV} communication: A
  connectivity-constrained trajectory optimization perspective,'' \emph{{IEEE}
  Trans. Commun.}, vol.~67, no.~3, pp. 2580--2604, Mar. 2019.

\bibitem{kosta2013interference}
C.~Kosta, B.~Hunt, A.~U. Quddus, and R.~Tafazolli, ``On interference avoidance
  through inter-cell interference coordination ({ICIC}) based on {OFDMA} mobile
  systems,'' \emph{{IEEE} Commun. Surveys Tuts.}, vol.~15, no.~3, pp. 973--995,
  3rd Quart. 2013.

\bibitem{hamza2013survey}
A.~S. Hamza, S.~S. Khalifa, H.~S. Hamza, and K.~Elsayed, ``A survey on
  inter-cell interference coordination techniques in {OFDMA}-based cellular
  networks,'' \emph{{IEEE} Commun. Surveys Tuts.}, vol.~15, no.~4, pp.
  1642--1670, 4th Quart. 2013.

\bibitem{lin2018mobile}
\BIBentryALTinterwordspacing
X.~Lin \emph{et~al.}, ``Mobile networks connected drones: Field trials,
  simulations, and design insights.'' [Online]. Available:
  \url{https://arxiv.org/ftp/arxiv/papers/1801/1801.10508.pdf}
\BIBentrySTDinterwordspacing

\bibitem{yaj2018interference}
V.~Yajnanarayana, Y.-P.~E. Wang, S.~Gao, S.~Muruganathan, and X.~Lin,
  ``Interference mitigation methods for unmanned aerial vehicles served by
  cellular networks,'' in \emph{Proc. IEEE 5G World Forum (5GWF)}, Silicon
  Valley, CA, USA, Jul. 2018.

\bibitem{amorim2018measured}
R.~Amorim \emph{et~al.}, ``Measured uplink interference caused by aerial
  vehicles in {LTE} cellular networks,'' \emph{{IEEE} Wireless Commun. Lett.},
  vol.~7, no.~6, pp. 958--961, Dec. 2018.

\bibitem{3GPP38901}
\BIBentryALTinterwordspacing
{3GPP-TR-38.901}, ``Study on channel model for frequencies from 0.5 to 100
  {G}hz,'' 2017, 3GPP technical report. [Online]. Available:
  \url{www.3gpp.org/DynaReport/38901.htm}
\BIBentrySTDinterwordspacing

\bibitem{beck2010sequential}
A.~Beck, A.~Ben-Tal, and L.~Tetruashvili, ``A sequential parametric convex
  approximation method with applications to nonconvex truss topology design
  problems,'' \emph{J. GlobalOpt}, vol.~47, no.~1, pp. 29--51, 2010.

\bibitem{boyd2009convex}
S.~Boyd and L.~Vandenberghe, \emph{Convex optimization}.\hskip 1em plus 0.5em
  minus 0.4em\relax Cambridge, UK: Cambridge university press, 2009.

\bibitem{phuong2003unified}
N.~T.~H. Phuong and H.~Tuy, ``A unified monotonic approach to generalized
  linear fractional programming,'' \emph{J. Global Optimization}, vol.~26,
  no.~3, pp. 229--259, Jul. 2003.

\bibitem{dahlman20134g}
E.~Dahlman, S.~Parkvall, and J.~Skold, \emph{{4G: LTE/LTE-advanced} for mobile
  broadband}.\hskip 1em plus 0.5em minus 0.4em\relax Oxford, UK: Academic
  press, 2013.

\bibitem{3GPP36819}
\BIBentryALTinterwordspacing
{3GPP-TR-36.819}, ``Coordinated multi-point operation for {LTE} physical layer
  aspects,'' 2013, 3GPP technical report. [Online]. Available:
  \url{www.3gpp.org/dynaReport/36819.htm}
\BIBentrySTDinterwordspacing

\bibitem{3GPP36423}
\BIBentryALTinterwordspacing
{3GPP-TR-36.423}, ``{X2} application protocol ({X2AP}),'' 2019, 3GPP technical
  report. [Online]. Available: \url{www.3gpp.org/dynareport/36423.htm}
\BIBentrySTDinterwordspacing

\bibitem{ballanis2016antenna}
C.~A. Ballanis, \emph{Antenna theory analysis and design}.\hskip 1em plus 0.5em
  minus 0.4em\relax New York, USA: John Willey and Son's Inc., 2016.

\bibitem{liu2012achieving}
L.~Liu, R.~Zhang, and K.-C. Chua, ``Achieving global optimality for weighted
  sum-rate maximization in the {K}-user {G}aussian interference channel with
  multiple antennas,'' \emph{{IEEE} Trans. Wireless Commun.}, vol.~11, no.~5,
  pp. 1933--1945, May 2012.

\end{thebibliography}
\bibliographystyle{IEEEtran}

\end{document}